\documentclass[prapplied,reprint,showpacs,superscriptaddress]{revtex4-2}
\usepackage{color}
\usepackage{amsfonts}
\usepackage{amsmath}
\usepackage{autobreak}
\usepackage{graphicx}
\usepackage{textcomp}
\usepackage{hyperref}
\usepackage[lmargin=2.1cm, rmargin=2.1cm,tmargin=2.97cm,bmargin=2.97cm]{geometry}
\usepackage{graphicx}
\usepackage{geometry}
\usepackage{siunitx}
\usepackage{mathbbol}
\usepackage{color}
\usepackage{ulem}
\usepackage{xcolor}
\graphicspath{ {./images/} }
\usepackage{amssymb}
\begin{document}

\title{Quantum integrated sensing and communication via entanglement}

\author{Yu-Chen Liu}
\affiliation{State Key Laboratory of Low-Dimensional Quantum Physics and Department of Physics, Tsinghua University, Beijing 100084, China}
\author{Yuan-Bin Cheng}
\affiliation{State Key Laboratory of Low-Dimensional Quantum Physics and Department of Physics, Tsinghua University, Beijing 100084, China}
\author{Xing-Bo Pan}
\affiliation{State Key Laboratory of Low-Dimensional Quantum Physics and Department of Physics, Tsinghua University, Beijing 100084, China}
\author{Ze-Zhou Sun}
\affiliation{State Key Laboratory of Low-Dimensional Quantum Physics and Department of Physics, Tsinghua University, Beijing 100084, China}
\author{Dong Pan}
\email{pandong@baqis.ac.cn}
\affiliation{Beijing Academy of Quantum Information Sciences, Beijing 100193, China}
\author{Gui-Lu Long}
\email{gllong@mail.tsinghua.edu.cn}
\affiliation{State Key Laboratory of Low-Dimensional Quantum Physics and Department of Physics, Tsinghua University, Beijing 100084, China}
\affiliation{Beijing Academy of Quantum Information Sciences, Beijing 100193, China}
\affiliation{Beijing National Research Center for Information Science and Technology, Beijing 100084, China}
\affiliation{Frontier Science Center for Quantum Information, Beijing 100084, China}

\date{\today}
\begin{abstract}
Quantum communication and quantum metrology are widely compelling applications in the field of quantum information science, and quantum remote sensing is an intersection of both. Despite their differences, there are notable commonalities between quantum communication and quantum remote sensing, as they achieve their functionalities through the transmission of quantum states. Here we propose a novel quantum integrated sensing and communication (QISAC) protocol, which achieves quantum sensing under the Heisenberg limit while simultaneously enabling quantum secure communication through the transmission of entanglements. We have theoretically proven its security against eavesdroppers. The security of QISAC is characterized by the secrecy capacity for information bit as well as asymmetric Fisher information gain for sensing. Through simulations conducted under the constraints of limited entanglement resources, we illustrate that QISAC maintains high accuracy in the estimation of phase. Hence our QISAC offers a fresh perspective for the applications of future quantum networks.
    
\end{abstract}
\maketitle

\section{Introduction}
\label{sec:intro}
Information science has embarked on a revolutionary step with the integration of quantum properties, where quantum secure communication and metrology are major breakthroughs of quantum information science. Relying on quantum entanglement as a core resource~\cite{horodecki2009quantum}, cryptography or communication paradigms with information-theoretic security, such as quantum key distribution~\cite{ekert1991quantum}, quantum secret sharing~\cite{hillery1999quantum,xiao2004efficient}, and quantum secure direct communication (QSDC)~\cite{long2002theoretically}, have been established. QSDC enables secure and reliable transmission of private messages in quantum channels susceptible to eavesdropping and noise~\cite{deng2003two,shapiro2019quantum,chandra2021direct,wu2022quantum,pan2023evolution,sternberg2024secure,pan2023free}, with high-capacity features~\cite{cao2023realization,patra2023dimensional,sephton2023quantum}. Leveraging the violation of the Bell-inequality, it achieves device-independent security, meaning secure communication can be achieved without trusting the inner workings of communication devices~\cite{zhou2020device}. Experimental results have shown that they can connect legitimate users to achieve on-demand information sharing~\cite{zhang2017quantum,zhu2017experimental,nirala2023information} and realize a multi-user quantum information network~\cite{qi202115}.

Quantum sensing~\cite{degen2017quantum} generally refers to the utilization of quantum states to detect and perceive variations in the probability distribution of a physical quantity. Quantum metrology~\cite{giovannetti2006quantum,giovannetti2011advances,toth2014quantum,pezze2018quantum,moore2023secure,zhang2021color} is a practical application of quantum sensing, which involves utilizing quantum systems for parameter estimation. The entanglement and superposition of quantum states have been demonstrated to enhance the sensitivity of parameter estimation, especially the Ramsey interferometry measurement~\cite{Ramsey1950A}, which can be used to measure the variations in physical quantities such as magnetic fields, gravity, and acceleration. Theoretically, a quantum entangled sensor has the potential to surpass the standard quantum limit (SQL) of any classical sensor, and to achieve measurement precision approaching the Heisenberg limit, where the variance of the estimated parameter decreases from proportional to $1/N$ to proportional to $1/N^{2}$. Although most quantum metrology protocols rely on multipartite entangled states, such as the NOON states, the constraints posed by quantum hardware in the noisy intermediate-scale quantum (NISQ) era~\cite{lau2022nisq} limit the practical applicability of these protocols, thereby increasing resource requirements.

A suitable alternative is the bipartite entangled state, such as the Bell state, which is the quantum state adopted by most quantum remote sensing (QRS) protocols~\cite{yin2020experimental,rahim2023quantum}. Similar to blind quantum computing achieved through the integration of quantum computing and quantum cryptography, the intersection of quantum cryptography and quantum metrology has led to the emergence of QRS. The concept of a quantum sensor network is gaining momentum in quantum metrology, with quantum cryptography playing a significant role in protecting sensing data during transmission. QRS protocols employ random quantum states to enhance the security of sensing parameters between separate remote nodes, whose security level is determined by the probability of detecting Eve's cheating. Through QRS schemes, one can achieve precise sensing of a remote system beyond the standard quantum limit while simultaneously ensuring the transmission of sensing parameters with quantum-native security. Specifically, users can implement a secure QRS scheme by sharing photon pairs in Bell states between local and remote sites~\cite{yin2020experimental}, where the security level is characterized by asymmetric Fisher information gain. Most recently, a quantum secure metrology protocol was constructed using bipartite entanglement pairs to provide high security and discussed the precision-security trade-off in a scenario with finite entanglement sources~\cite{rahim2023quantum}.

If taking a unified perspective on communication and sensing, we might integrate communication and sensing functionalities into a single system. This dual functionality can utilize resources more effectively. The introduction of quantum properties will further enhance the security of communication and sensing, as well as the precision of sensing. Finally, we are presented with a unique opportunity to develop a novel information processing infrastructure.

Against this background, we propose a quantum integrated sensing and communication (QISAC) protocol that implements simultaneous quantum sensing and QSDC using the same entangled pairs, rather than constructing separate sensing and communication systems based on two entangled pairs, which differs from previous protocol of Rahim \textit{et al.}~\cite{rahim2023quantum} that is focused exclusively on quantum-secure metrology. The QISAC protocol involves three primary participants: Alice, Bob, and Eve, with Alice and Bob being legitimate users and Eve being a potential eavesdropper. In such a protocol, the sensing component maintains superior sensing capability, which transcends the standard quantum limit and attains a measurement precision approaching the Heisenberg limit. Both the security of remote sensing and that of communication are maintained under collective attacks by Eve.

The paper is organized as follows. In Sec.~\ref{sec:pre}, we present the modified entanglement-based QSDC protocol and fundamental concepts in quantum metrology. In Sec.~\ref{sec:protocol}, we detail the proposed QISAC scheme, demonstrating that the protocol can achieve the Heisenberg limit in parameter sensing. In Sec.~\ref{sec:security}, we provide a security proof for the proposed QISAC protocol from both quantum sensing and quantum communication perspectives. Based on the physical structure, we simulate the performance of QISAC in Sec.~\ref{sec:performance}. Finally, we draw conclusions and provide an outlook in Sec.~\ref{sec:conclusion}.

\section{Elements}
\label{sec:pre}
\subsection{Modified two-step QSDC protocol} \label{sec:twostep}
\begin{figure}[bht]
  \centering
  \includegraphics[width=\linewidth]{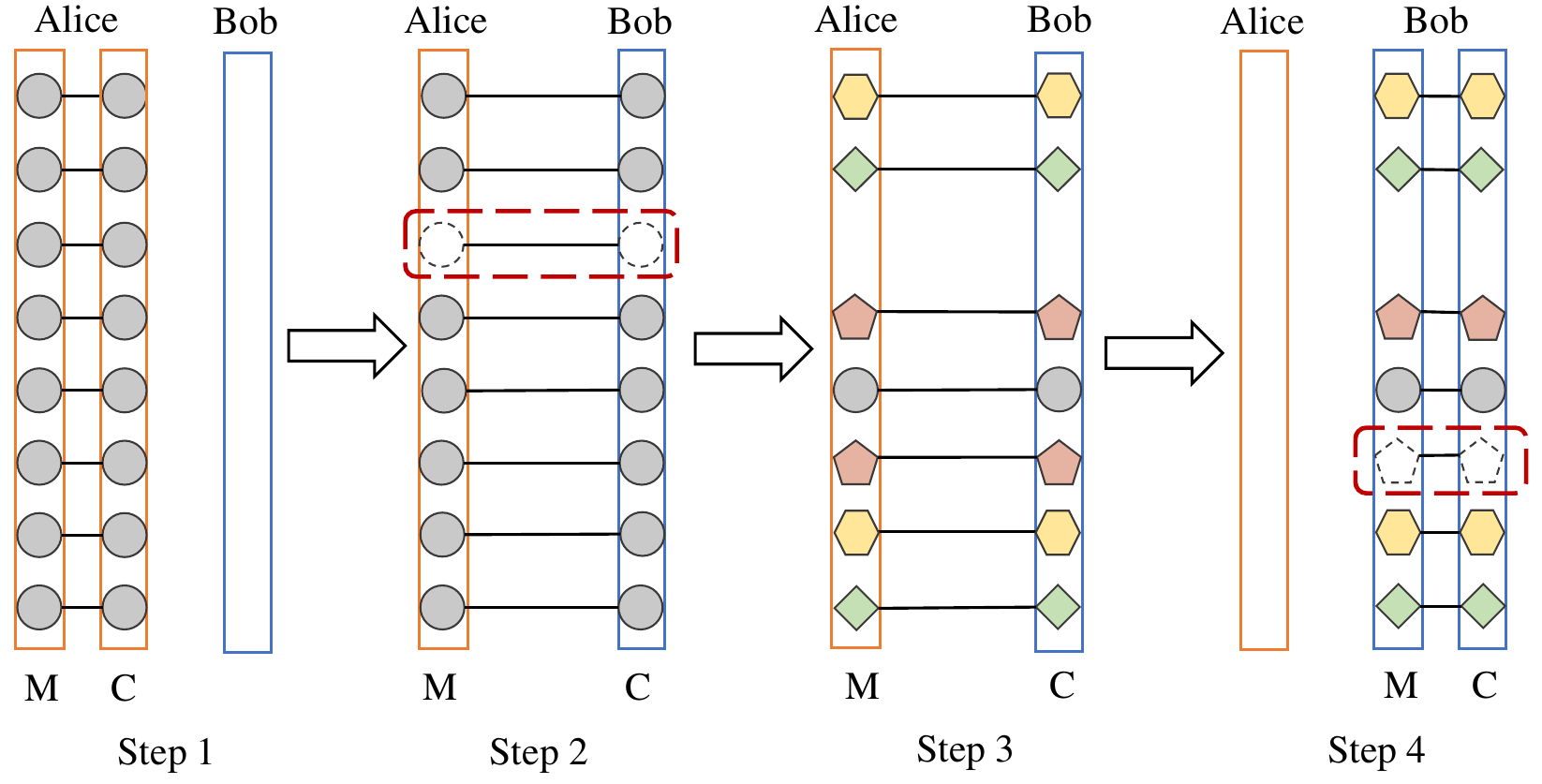}
  \caption{Schematic illustration of the modified two-step QSDC protocol. The circles, hexagons, rhombuses and pentagons represent the Bell state $\lvert \psi^-\rangle$, $\lvert \psi^+\rangle$, $\lvert \phi^-\rangle$ and $\lvert \phi^+\rangle$, respectively. The dashed circles are used for eavesdropping, and denoted with red dashed boxes.}
  \label{fig1:twostep}
\end{figure}

In the two-step QSDC scheme~\cite{deng2003two}, legitimate users Alice and Bob employ the following Einstein-Podolsky-Rosen (EPR) states to achieve the secure transmission of private information:
\begin{equation}
  \label{eq:bell}
  \begin{aligned}
    &\left | \psi^{-}   \right \rangle =\frac{1}{\sqrt{2} } \left ( \left | 0  \right \rangle_{M} \left | 1  \right \rangle_{C} -\left | 1  \right \rangle_{M} \left | 0  \right \rangle_{C}   \right ), \\
    &\left | \psi^{+}   \right \rangle =\frac{1}{\sqrt{2} } \left ( \left | 0  \right \rangle_{M} \left | 1  \right \rangle_{C} +\left | 1  \right \rangle_{M} \left | 0  \right \rangle_{C}   \right ),\\
    &\left | \phi^{-}   \right \rangle =\frac{1}{\sqrt{2} } \left ( \left | 0  \right \rangle_{M} \left | 0  \right \rangle_{C} -\left | 1  \right \rangle_{M} \left | 1  \right \rangle_{C}   \right ),\\
    &\left | \phi^{+}   \right \rangle =\frac{1}{\sqrt{2} } \left ( \left | 0  \right \rangle_{M} \left | 0  \right \rangle_{C} +\left | 1  \right \rangle_{M} \left | 1  \right \rangle_{C}   \right ).
  \end{aligned}
\end{equation}
They establish a consensus on the particular mapping of the $\left | \psi^{-} \right \rangle$, $\left | \psi^{+} \right \rangle$, $\left | \phi^{-} \right \rangle$, and $\left | \phi^{+} \right \rangle$ states to the classical two-bit classical information 00, 01, 10, and 11, respectively. By enhancing the eavesdropping detection part, the modified two-step protocol can resist malicious party~\cite{fahmi2008comment} securely. As shown in Fig.~\ref{fig1:twostep}, the detailed steps of this entanglement-based QSDC protocol are as follows.

\textit{Step 1, states preparation}. As shown in Fig.~\ref{fig1:twostep}, Alice prepares an ordered sequence of EPR pairs in state $\left | \psi^{-} \right \rangle$ and divides the sequence into two particle sequences. One of them is called the checking sequence or simply the C sequence. The remaining EPR partner particles consist of the other remaining sequence which is called the M sequence for short.

\textit{Step 2, qubits transmission and first eavesdropping detection.} The C sequence is sent from Alice to Bob. Bob randomly selects some particles from the C sequence and measures them using the bases $\sigma _x$, $\sigma _y$ or $\sigma_z$ randomly. Bob informs Alice of the positions and measurement bases of the chosen particles. Alice performs the same measurements on the corresponding partner particles in the M sequence and checks the results with Bob. They calculate the quantum bit error rate (QBER) and determine that no eavesdropper is interfering with the quantum channel, if the QBER is below the tolerance threshold. Otherwise, they terminate their communication. As depicted in Fig.~\ref{fig1:twostep}, the third EPR pair serves the purpose of the first eavesdropping detection.

\textit{Step 3, encoding.} After the first detection, Alice and Bob discard the particles in sequences M and C that were used for the eavesdropping detection. Then, Alice applies one of the following four unitary operators to each remaining particle in the M sequence to encode 00, 01, 10, or 11, respectively,
\begin{equation}
\label{eq:fouroperation}
  \begin{aligned}
    &I=| 0  \rangle  \langle 0 | + | 1  \rangle  \langle 1 | , \\
    &\sigma_z= | 0  \rangle  \langle 0 | - | 1  \rangle \langle 1  |,\\
    &\sigma_x= | 0 \rangle  \langle 1  | + | 1  \rangle  \langle 0 |,\\
    &i\sigma_y= | 0 \rangle  \langle 1  | - | 1  \rangle  \langle 0 |.
  \end{aligned}
\end{equation}
Particularly, Alice has to apply a small trick within the M sequence for the second eavesdropping detection. She randomly selects a small subgroup of particles from the remaining M sequence as samples and performs one of the four unitary operators on them randomly. These states carry random numbers that are used in the second detection.

\textit{Step 4, measurement and the second eavesdropping detection.} Upon receiving the M sequence, Bob combines it with the corresponding particle in the C sequence and performs the Bell-state measurement. At this stage, the particles of the C sequence remain one-to-one correlated with those of the M sequence ideally. Alice informs Bob about the positions of the sampling states and the specific unitary operators applied. By checking the chosen sampling pairs, Bob estimates the QBER of the whole EPR sequence. The sixth EPR pair with the dashed line is encoded with the random bits 11 for eavesdropping detection, as shown in Fig.~\ref{fig1:twostep}. If the QBER is reasonably low, Alice and Bob can trust the process. If not, they will abandon the transmission and revert to step 1. Ultimately, Bob retrieves the message bits transmitted by Alice.

Considering the depolarization channel, a fundamental and highly important quantum channel that is extremely useful for modeling noise in quantum communication~\cite{beaudry2013security,wu2019security}, we derive a lower bound for the secrecy capacity for the modified two-step QSDC protocol, based on the work in Ref.~\cite{wu2019security}. Specifically, we have
\begin{equation}
  \label{eq:twostep}
  C_s\ge Q_B[2-h_4(\textbf{e})]-Q_E[2h(\varepsilon)],
\end{equation}
where $h_4(\cdot)$ is the four-array Shannon entropy, $h(\cdot)$ is the Shannon entropy, $Q_B$ and $Q_E$ are the reception rates of Bob and Eve, respectively, $\textbf{e}$ is the error rate distribution of the main channel, and $\varepsilon$ is the QBER obtained in step 2. 

\subsection{Quantum metrology} \label{sec:qmtheory} 
\begin{figure}[bht]
  \centering
  \includegraphics[width=\linewidth]{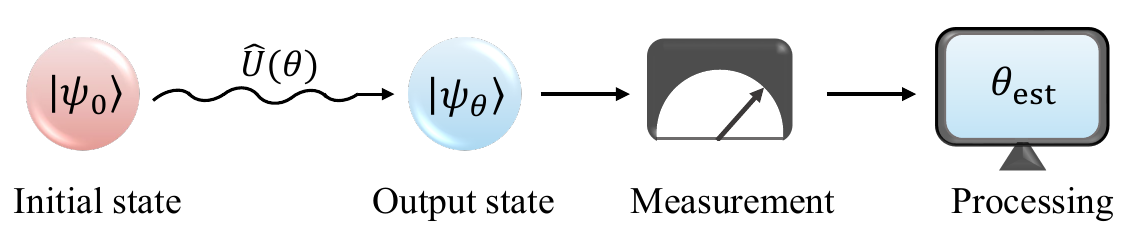}
  \caption{The general process of quantum metrology.} 
  \label{fig2:qm}
\end{figure}

In the field of quantum metrology, a fundamental task is to estimate small parameters by measuring the expectation value of a Hermitian measurement operator~\cite{toth2014quantum,zhong2014optimal}, which we will elaborate on below. Generally, quantum single parameter estimation can be abstractly modeled in four steps, as illustrated in Fig.~\ref{fig2:qm}: ({\romannumeral1}) preparing the initial probe states $\rho_0$, ({\romannumeral2}) parameterizing the initial states under the dynamical evolution of the parameter-dependent system, ({\romannumeral3}) performing the appropriate observable measurement on the output probe states, and ({\romannumeral4}) processing the measurement results to formulate an estimation of the parameter $\theta $. In step ({\romannumeral2}), the unitary parameterization is expressed as $U_\theta =e^{-iH\theta }$, where $H$ is the Hamiltonian of the system and $\theta $ is the parameter to be estimated.

Next, we describe the concept of Fisher information in quantum metrology. Fisher information quantifies the information conveyed by a set of observable random variables about the parameters awaiting measurement. Specifically, we assume that parameter estimation of $\theta $ relies on the observable $\hat{O}$. According to the Born rule, the conditional probability $p(x | \theta)$, alternatively termed the likelihood function $L(\theta |x)$, is defined as follows,
\begin{equation}
  L(\theta|x)=P(x | \theta)=Tr[\rho(\theta) \hat{O}(x)],
\end{equation}
where $\hat{O}(x)$ corresponds to the projector corresponding to the measured value $x$. By performing repeated measurements on the same system, the likelihood function for the parameters to be estimated can be expressed as the product of the likelihood functions for individual measurements, namely
\begin{equation}
  \label{eq:likelihood}
  L(\theta|x)= {\textstyle\prod_{i=1}^{m} p_i(x_i | \theta)}.
\end{equation}
Usually, we use the log-likelihood function
\begin{equation*}
 L_{\rm log} (\theta|x)={\rm ln}L(\theta|x)=\sum_{i=1}^{m}{\rm ln} p_i(x_i | \theta).
\end{equation*}
The distribution function naturally satisfies the normalization condition $\sum_{i=1}^{m}p_i(x_i | \theta)=1$. Furthermore, the associated classical Fisher information (CFI) is
\begin{equation}
 \label{eq:cfi}
\begin{aligned}
  F(\theta)&=\left \langle \left ( \frac{\partial L_{log}(x|\theta)}{\partial \theta}  \right ) ^2 \right \rangle _\theta\\
  &=\sum_{i}\frac{1}{P(x_i|\theta)}  \left ( \frac{\partial P(x_i|\theta)}{\partial\theta}  \right ) ^2.
\end{aligned}
\end{equation}
Observing Eq.~(\ref{eq:cfi}), it is evident that different measurement operators lead to varying CFI values. The quantum Fisher information (QFI), which corresponds to the CFI of an optimal observable, serves as the maximum estimation precision~\cite{zhong2014optimal}. It is denoted by
\begin{equation}
  \label{eq:qfi}
  F_q={\rm Tr}(\rho_\theta \hat{L} ^2),
\end{equation}
where the Hermitian operator $\hat{L}$ is the symmetric logarithmic derivative~\cite{braunstein1996generalized} and determined by
\begin{equation}
  \partial_\theta\rho=\frac{1}{2}(\rho \hat{L}+\hat{L} \rho).
\end{equation}
The Fisher information directly correlates with the perturbation induced by even minor parameter fluctuations of probe states. However, pinpointing the optimal operator for achieving the highest estimation precision can be challenging through mere trial and error. Fortunately, it is feasible to establish an upper bound on parameter estimation precision applicable to any operator selection. Irrespective of the chosen measurement approach, the quantum Cram\'{e}r-Rao bound provides an ultimate limit on the precision of unbiased estimate~\cite{toth2014quantum}, which is expressed as~\cite{helstrom1969quantum}
\begin{equation}
  \delta^2 \theta\ge \frac{1}{\nu F},
\end{equation}
where $\nu$ represents the number of measurement repetitions and $F$ denotes the QFI. From estimation theory, the estimation precision $\delta^2\theta_{\text{est}}$ is measured by the units-correlated mean-square deviation of the estimator $\theta_{\text{est}}$ from the true value $\theta$~\cite{zhong2014optimal}, namely, 
\begin{equation}
  \delta^2\theta_{\text{est}}:= {\left \langle \left ( \frac{\theta_{\text{est}}}{\left | \partial_\theta \left \langle \theta_{\text{est}} \right \rangle_{\text{av}} \right | -\theta}  \right ) ^2 \right \rangle}_{\text{av}},
\end{equation}
where the bracket $\langle\cdot\rangle_{\text{av}}$ denotes the statistical average, and the derivative $\partial_\theta \left \langle \theta_{\text{est}} \right \rangle\equiv \partial \left \langle \theta_{\text{est}} \right \rangle /\partial \theta$ accounts for the difference in the units of $\theta_{\text{est}}$ and $\theta$.

On the other hand, error-propagation theory is widely accepted in experimental settings~\cite{toth2014quantum}. According to this theory, the variance of error-propagation $\delta^2\theta_{\text{ep}}$ reduces to the expectation value of an observable $\hat{O}$ after $\nu$ repetitions, yielding
\begin{equation}
  \label{eq:propagation}
  \delta^2\theta_{\text{ep}}=\frac{\left \langle \Delta \hat{O}^2 \right \rangle }{\nu \left | \partial_\theta \left \langle \hat{O} \right \rangle  \right|^2 } ,
\end{equation}
where $\left \langle \Delta \hat{O}^2 \right \rangle =\left \langle \hat{O}^2\right \rangle -\left \langle \hat{O} \right \rangle^2$. In general, the relationship between the two types of estimation can be expressed as an inequality \cite{zhong2014optimal}: $\delta^2\theta_{\text{est}}\ge \delta^2\theta_{\text{ep}} \ge (\nu F)^{-1}$. For sufficient large $\nu$, using optimal observable $\hat{O}_{\text{opt}}$ enables both $\delta^2\theta_{\text{est}}$ and $\delta^2\theta_{\text{ep}}$ reach the same quantum Cram\'{e}r-Rao bound, \textit{i.e.}, the saturation of the former implies that of the latter. It is also a necessary and sufficient condition, which means that when $\delta^2 \theta_{ep}$ reaches the  quantum Cram\'{e}r-Rao bound saturation, the operator $\hat{O}$ can be determined as an optimal observable $\hat{O}_{\text{opt}}$.

\section{Quantum integrated sensing and communication protocol}
\label{sec:protocol}
\subsection{Protocol description and precision calculation}
\begin{figure*}[bht]
  \centering
  \includegraphics[width=\linewidth]{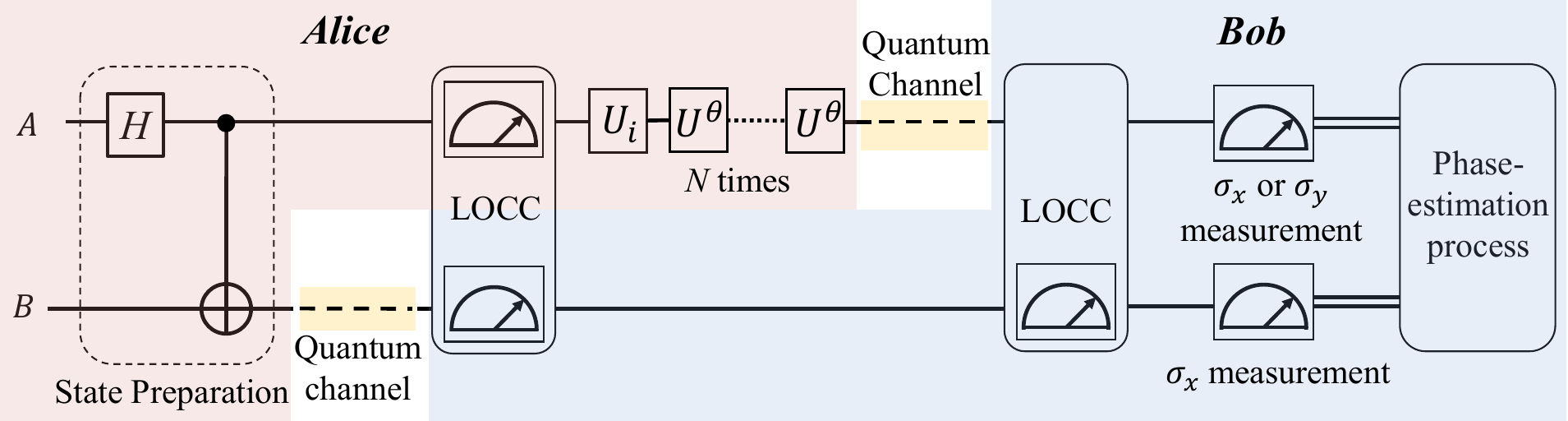}
  \caption{Schematic quantum circuit of the QISAC protocol, where the pink area and the blue area represent Alice and Bob, respectively.}
  \label{fig3:protocol}
\end{figure*}

In this section, we describe the QISAC protocol in detail. We consider a scenario where Alice holds a collection of confidential messages that she intends to transmit to Bob, meanwhile Bob aims to conduct remote measurements at Alice's location without compromising sensitive information to Eve. To achieve this goal, they harness the entangled photons detailed in Eq.~(\ref{eq:bell}) as quantum state resources for implementing the protocol. As depicted in Fig~\ref{fig3:protocol}, the QISAC protocol consists of the following six steps.

Prior to the communication, Alice and Bob agree that the state $|\psi^{-}\rangle$ corresponds to a specific mapping of bit 0, while the state $|\phi^{-}\rangle$ corresponds to a specific mapping of bit 1. Here, we denote the ratio of EPR pairs carrying the confidential information within the initial EPR pairs as $p_e\in (0,1)$, while the remaining ratio $1-p_e$ of EPR pairs serves for eavesdropping detection. We will discuss the trade-off related to the choice of $p_e$ under finite resource constraint in Sec.~\ref{sec:limited}.

\textit{Step 1, states preparation and distribution.}
Alice generates $m$ EPR pairs $\lvert\psi^-\rangle^{\otimes m}$ as the initial probe states, denoted by $(\text{A}_1 \text{B}_1, \text{A}_2 \text{B}_2, \cdots, \text{A}_i \text{B}_i, \cdots,\text{A}_m \text{B}_m)$. Here, the subscript $i$ represents the order index of EPR pairs in the sequence, while A and B respectively denote a photon and its partner in the EPR pairs. Alice divides the initial EPR pairs into two partner photon sequences, referred to as sequence $S_{\text{A}}$ and sequence $S_{\text{B}}$. She keeps $S_{\text{A}}$ for herself and sends $S_{\text{B}}$ to Bob via the quantum channel. This step completes the initial distribution of entangled pairs.

\textit{Step 2, first eavesdropping detection.} After Bob receives the photon sequence $S_B$, Alice and Bob follow these steps to detect eavesdropping: ({\romannumeral1}) Bob randomly selects photons from sequence $S_B$ in positions corresponding to a proportion $(1-p_e)/2$ of the initial number of EPR pairs and measures the selected photons randomly on the $\sigma_x$, $\sigma_y$, or $\sigma_z$ basis, ({\romannumeral2}) Bob informs Alice about the positions and measurement bases of the selected photons and shares the measurement outcomes with her through a classical authenticated channel, ({\romannumeral3}) Alice measures the corresponding EPR partner photons in $S_A$, using the same measurement bases chosen by Bob, and then compares the measurement results. Alice then obtains the opposite results compared to Bob, provided that no eavesdropper contaminates the quantum channel. If the error rate is below the tolerance threshold, Alice and Bob conclude that no eavesdropper was present during the first transmission and proceed to the next step. Otherwise, they terminate the protocol.

\textit{Step 3, encoding and sensing.} In sequence $S_A$, the remaining photons serve to carry confidential messages and the parameter to be estimated. Alice first encodes the plaintext by applying
\begin{equation}
U_0=I=| 0  \rangle  \langle 0 | + | 1  \rangle  \langle 1 |,
\end{equation} 
or
\begin{equation}
U_1=\sigma_x=| 0  \rangle  \langle 1 | + | 1  \rangle  \langle 0 |,
\end{equation} 
where $U_0$ and $U_1$ map the bit $0$ and $1$ on to quantum state, respectively. Then, she employs these information-encoded quantum states as initial probes to perform a sensing task $N$ times on the unknown system. The evolution operator of the system is expressed as
\begin{equation}
  U^\theta = e^{-iJ_z \theta}=| 0  \rangle  \langle 0 | +e^{i\theta } | 1  \rangle  \langle 1 |,
\end{equation}
where $\theta$ is the parameter to be estimated. Repeating the sensing system $N$ times can be understood as a phase accumulation process, which is related to the Ramsey interferometry measurement~\cite{Ramsey1950A} in quantum sensing. This parameterization process turns the encoded states into
\begin{equation}
  \begin{aligned}
\lvert \psi^{-}_{\theta}\rangle=\frac{1}{\sqrt{2}}(\lvert01\rangle-e^{iN\theta }\lvert10\rangle)
  \end{aligned}
\end{equation} 
and 
\begin{equation}
  \begin{aligned}
\lvert \phi^{-}_{\theta}\rangle=\frac{1}{\sqrt{2}}(\lvert00\rangle-e^{iN\theta }\lvert11\rangle).
  \end{aligned}
\end{equation} 
During the information encoding process, Alice applies an encoding trick within the sequence $S_A$ to guard against second eavesdropping. She randomly selects photons from sequence $S_A$ correct to $(1-p_e)/2$ of the initial EPR pairs as samples to undergo one of the four unitary operators described by Eq.~(\ref{eq:fouroperation}) in a disorderly manner. Importantly, these photons carry only random numbers for eavesdropping detection, and they are not used for sensing tasks. Alice keeps the positions of sampling photons secret until confirming that Bob has received the sequence $S_A$. 

\textit{Step 4, transmission and second eavesdropping detection.} Alice sends the encoded EPR partner photon sequence $S_A$ to Bob via the quantum channel. Upon receiving the photon sequence, Bob is informed of the positions of the samples and the specific unitary operators applied to them. To ensure communication security, Bob performs Bell-state measurement on these photons and estimates the QBER. If the QBER of the checking pairs is reasonably low, Alice and Bob can trust the process and proceed to decode plaintext and estimate $\theta$. Otherwise, they will abort and return to step 1.

Interestingly, Eve can perturb the qubits using a man-in-the-middle attack during the second transmission but cannot steal the confidential messages. This limitation arises because one cannot read information from a single particle of an EPR pair alone, which results in the maximum mixed state for an eavesdropper. Furthermore, any perturbation introduced by Eve can be detected based in the second detection.

The following steps 5 and 6 involve reading out the confidential messages and estimating the parameter $\theta$.

\textit{Step 5, measurement.} Bob measures each encoded EPR pair $(\text{A}_\textit{i}\text{B}_\textit{i})$ on two measuring bases randomly. He randomly chooses to perform the observable $\hat{O_1}$ or $\hat{O_2}$ with probabilities $p_o$ and $1-p_o$, respectively, where 
\begin{equation}
  \hat{O_1}=\sigma_x\otimes \sigma_x=
  \begin{pmatrix}
    0& 0 & 0 & 1\\
    0& 0 & 1 & 0\\
    0& 1 & 0 & 0\\
    1& 0 & 0 &0
  \end{pmatrix}
\end{equation}
and
\begin{equation}
  \hat{O_2}=\sigma_y\otimes \sigma_x=
  \begin{pmatrix}
    0& 0 & 0 & -i\\
    0& 0 & -i & 0\\
    0& i & 0 & 0\\
    i& 0 & 0 &0
  \end{pmatrix}
  .
\end{equation}

\textit{Step 6, the parameter estimation process.}
After measurement, Bob decodes the information bit and estimates the parameter using the raw measurement data. The operators $\hat{O_1}$ and $\hat{O_2}$ each have four eigenvectors, corresponding to four detectors. As illustrated in Table~\ref{tab:tab1}, we calculate the probability distribution for different responses with the two observables, where
\begin{equation}
\begin{aligned} 
\lvert t_1\rangle&=-\lvert00\rangle+\lvert11\rangle,\\
\lvert t_2\rangle&=\lvert00\rangle+\lvert11\rangle,\\
\lvert t_3\rangle&=\lvert01\rangle+\lvert10\rangle,\\
\lvert t_4\rangle&=-\lvert01\rangle+\lvert10\rangle, 
\end{aligned} 
\end{equation}
are the eigenvectors of $\hat{O_1}$, and
\begin{equation}
\begin{aligned} 
\lvert r_1\rangle&=i\lvert00\rangle+\lvert11\rangle,\\
\lvert r_2\rangle&=-i\lvert00\rangle+\lvert11\rangle,\\
\lvert r_3\rangle&=i\lvert01\rangle+\lvert10\rangle,\\
\lvert r_4\rangle&=-i\lvert01\rangle+\lvert10\rangle,
\end{aligned} 
\end{equation}
are the eigenvectors of $\hat{O_2}$. Obviously, the eigenvectors of $\hat{O_1}$ are four Bell-states.

\begin{table*}[htb]
  \centering
    \caption{The specific probabilities of distinct quantum states for two observables.}
    \label{tab:tab1}
    \resizebox{\textwidth}{!}{
    {\renewcommand{\arraystretch}{1.4}
    \setlength{\tabcolsep}{2pt}
    \begin{tabular}{ccccc|cccc}
      \hline   \hline
      & \multicolumn{4}{c|}{$\hat{O_1}$} & \multicolumn{4}{c}{$\hat{O_2}$} \\ \hline
      Eigenvectors& $\lvert t_1\rangle$         & $\lvert t_2\rangle$          & $\lvert t_3\rangle$         & $\lvert t_4\rangle$  &     $\lvert r_1\rangle$        &     $\lvert r_2\rangle$        &     $\lvert r_3\rangle$       &     $\lvert r_4\rangle$       \\ \hline
      Eigenvalues&      -1      &       1     &      1      &      -1     &       -1     &      1      &      1    &      -1    \\ \hline
      $P_{\lvert \psi^{-}_{\theta}\rangle}$&      0      &       0     &        $\frac{1}{2}(1-\cos N\theta)$    &      $\frac{1}{2}(1+\cos N\theta)$     &       0     &        0    &       $\frac{1}{2}(1+\sin N\theta)$    &    $\frac{1}{2}(1-\sin N\theta)$       \\\hline
      $P_{\lvert \phi^{-}_{\theta}\rangle}$&     $\frac{1}{2}(1+\cos N\theta) $     &       $ \frac{1}{2}(1-\cos N\theta) $   &      0      &       0    &       $ \frac{1}{2}(1+\sin N\theta)$    &      $ \frac{1}{2}(1-\sin N\theta) $    &       0    &       0   \\
      \hline         \hline
    \end{tabular}}}
\end{table*}

Explicitly, this means that when detector 3 or 4 of $\hat{O_1}$ and detector 3 or 4 of $\hat{O_2}$ click, it corresponds to the quantum state $\lvert \psi^- \rangle$, representing bit 0. Conversely, when detector 1 or 2 of $\hat{O_1}$ and detector 1 or 2 of $\hat{O_2}$ click, it corresponds to the quantum state $\lvert \phi^- \rangle$, denoting bit 1. By analyzing the distribution of detector responses, we can calculate the expected values of the two observables and estimate the phase value. The expected values of $\hat{O_1}$ and $\hat{O_2}$ are
\begin{equation}
  \label{eq:expected}
  \begin{aligned}
    &\langle \hat{O_1} \rangle={\rm Tr}(\rho \hat{O_1})=-{\rm cos}N\theta,\\
    &\langle \hat{O_2} \rangle={\rm Tr}(\rho \hat{O_2})=-{\rm sin}N\theta,
  \end{aligned}
\end{equation}
where
\begin{equation}
  \label{eq:rho}
  \rho=\frac{1}{2}(\lvert \psi^{-}_{\theta}\rangle\langle \psi^{-}_{\theta}\rvert+\lvert \phi^{-}_{\theta}\rangle\langle \phi^{-}_{\theta}\rvert)
\end{equation}
refers to the density matrix of mixed states received by Bob. The experimental expected value is obtained by classical calculation of measurement data, with the formula $\langle \hat{O} \rangle= \sum_{i}x_ip_i$, where $x_i$ and $p_i$ are the eigenvalues and response ratios of the $i$-th detector. Then, Bob obtains an estimate $\theta$ from Eq.~(\ref{eq:expected}).

For each selected value of $\theta$, Bob theoretically calculates and obtains the same estimation value from two operators. However, the dependencies of $\theta$ on probabilities are complicated for the actual shared state used in experiment, even though the fidelity between the shared state is greater than 0.99~\cite{yin2020experimental}. To solve the problem of fluctuating values obtained by distinct observables, Bob separately calculates the expected values of two operators. Then he estimates the parameter by two expected values, specifically $\langle\hat{O_1}\rangle=-\text{cos}N\theta_1$ and $\langle\hat{O_2}\rangle=-\text{sin}N\theta_2$. The parameter estimate is then taken as the average of $\theta_1$ and $\theta_2$, given by
\begin{equation}
  \theta_{\text{est}}=\frac{1}{2}(\theta_1+\theta_2).
\end{equation}

After repeating the estimation measurement $\nu=p_em$ times, we can evaluate $\delta^2 \theta$ in error-propagation formula, Eq.~(\ref{eq:propagation}), using the expectations and variances of two observables. The variances of two observables are
\begin{equation}
  \begin{aligned}
    &\langle \Delta\hat{O_1}^2\rangle=1-{\rm cos}^2N\theta={\rm sin}^2N\theta,\\
    &\langle \Delta\hat{O_2}^2\rangle=1-{\rm sin}^2N\theta={\rm cos}^2N\theta.
  \end{aligned}
\end{equation}
Hence the variance $\delta^2 \theta_1$ and $\delta^2 \theta_2$ according to Eq.~(\ref{eq:propagation}) are
\begin{equation}
  \begin{aligned}
  \delta^2 \theta_1=\frac{1}{p_ep_omN^2},\\
  \delta^2 \theta_2=\frac{1}{p_e(1-p_o)mN^2}.
  \end{aligned}
\end{equation}
By using the error-propagation theory, the error on $\theta$, namely the variance $\delta^2 \theta$ can be obtained easily by
\begin{equation}
  \delta^2 \theta=\frac{\delta^2\theta_1+\delta^2\theta_2}{4}=\frac{1}{4p_ep_o(1-p_o)mN^2}.
\end{equation}
To minimize the variance, the $\hat{O_1}$ and $\hat{O_2}$ need to be performed with equal probability, hence we arrive
\begin{equation}
  p_o=1-p_o=\frac{1}{2},
\end{equation}
that is, Bob performs the observables $\hat{O_1}$ and $\hat{O_2}$ to measure the encoded EPR pairs randomly with 50\% probability. The variance $\delta^2 \theta$ from error-propagation theory becomes
\begin{equation}
  \label{eq:variance-ep}
  \delta^2 \theta_{\rm ep}=\frac{1}{p_emN^2}.
\end{equation}
The quantum Fisher information defined in Eq.~(\ref{eq:qfi}) can be expressed by~\cite{liu2014quantum}
\begin{equation}
  \begin{aligned}
  F_q=&\sum_{i=1}^{s} \frac{1}{p_i}(\partial_\theta p_i)^2+\sum_{i=1}^{s} 4p_i \langle \partial_\theta \psi_i  | \partial_\theta \psi_i  \rangle\\
  &-\sum_{i,j=1}^{s} \frac{8p_ip_j}{p_i+p_j}\lvert \langle \psi_i  | \partial_\theta \psi_j  \rangle\rvert^2,
  \end{aligned}
\end{equation}
where the spectral decomposition of the density matrix is given by $\rho_\theta= {\textstyle \sum_{i=1}^{s}} p_i\lvert\psi_i\rangle\langle\psi_i\rvert$. Furthermore, $s$ is the dimension of the support set of $\rho_\theta$. From this equation, one can find that the quantum Fisher information for a non-full rank density matrix is determined by its support. By plugging our system $\rho$ from Eq.~(\ref{eq:rho}) into the above equation, we obtain that the QFI is
\begin{equation}
  F_q(\rho)=N^2,
\end{equation}
and the corresponding quantum Cram\'{e}r-Rao bound is
\begin{equation}
  \delta^2\theta_{\rm est} \ge \frac{1}{p_emN^2},
\end{equation}
where the number of the measurement $\nu$ equals $p_em$. By comparing Eq.~\ref{eq:variance-ep} with the equation above, we find that the precision of parameter sensing can reach the quantum Cram\'{e}r-Rao bound. This implies that our approach saturates the ultimate limit of unbiased estimate in QISAC protocol.

It is easy to observe that our scheme saturates the parameter estimation precision up to the Heisenberg limit sensitivity ($\sim 1/N^2$). It is worth noting that, similar to the standard quantum limit, there is no unique definition of the Heisenberg limit~\cite{braun2018quantum}. This is because the definition depends on which resource is considered as the limiting factor. Here, we typically use the number of times of sensing the system to determine the precision limit, which is the approach taken in most quantum sensing studies~\cite{giovannetti2006quantum,zhong2014optimal,moore2023secure,rahim2023quantum}.

Recalling the QFI, which geometrically captures the rate of change of the evolved probe state under an infinitesimal variation of the parameter, it represents the maximum CFIs and signifies the optimal observable, denoted as $F_q={\rm sup}_{\hat{O}}F^{\hat{O}}$. By substituting the probabilities $P(x_i|\theta)$ in Eq.~(\ref{eq:cfi}) with the values in Table~\ref{tab:tab1}, we obtain the CFIs for $\hat{O_1}$ and $\hat{O_2}$, namely,
\begin{equation}
  \begin{aligned}
&F^{\hat{O_1}}=N^2,\\
&F^{\hat{O_2}}=N^2.
  \end{aligned}
\end{equation}
This means that both $\hat{O_1}$ and $\hat{O_2}$ are optimal observables because the classical Fisher information saturates the quantum Fisher information.

In our QISAC protocol, Alice only needs to perform single-qubit measurements, eliminating the need for entangling operations. This aspect is crucial for reducing technological requirements. Moreover, our scheme is capable of conducting separable measurements, surpassing the standard quantum limit ($1/N$), and achieving Heisenberg-limit sensitivity ($1/N^2$). These features support the conclusion demonstrated in~\cite{zhong2014optimal}. Considering the operational complexity, our protocol offers practicality.

\subsection{Further refinement of QISAC protocol}
In terms of trading physical resources for time, achieving the same precision as the Heisenberg limit ($1/N^2$) is possible classically by repeatedly sensing the same system $N$ times. This strategy is analogous to utilizing the NOON state~\cite{giovannetti2006quantum}, which takes the form of
\begin{equation}
  \lvert \psi\rangle=\frac{1}{\sqrt{2}}\left(\lvert 0\rangle ^{\otimes N}+\lvert 1\rangle ^{\otimes N}\right).
\end{equation}
The fundamental concept of enhancing precision through a multi-round strategy is intuitive. By sensing $N$ passes, it effectively replaces $\theta$ with $N\theta$ in both the probe states and the corresponding detection probabilities. Consequently, the CFI undergoes a quadratic enhancement, scaling by a factor of $N^2$ as depicted in Eq.~(\ref{eq:cfi}). Similarly, the variance propagated from measurement errors also experiences an enhancement, scaling by a factor of $N^2$ as shown in Eq.~(\ref{eq:propagation}).

Due to the challenging preparation and susceptibility to decoherence of NOON states, we opt for the multi-round strategy, where the same photon in sequence $S_\text{A}$ passes the unknown system $N$ times. Although this approach offers a straightforward and convenient means to improve measurement precision, it introduces a new challenge, as it narrows the estimable range of the parameter to $[0,2\pi/N)$. Expanding the parameter range to $[0,2\pi)$ is the problem we need to address.

Various approaches exist to address this issue~\cite{berry2009perform}. Alice could declare the range within which the transmitted parameter lies, but this would also narrow the range of parameter for Eve. Instead, we set different numbers of passes, $N$. Specifically, a small fraction of photons in sequence $S_A$ are randomly selected to sense the unknown system only once, while the remaining photons sense the system $N$ times. Once both parties confirm the security of communication, Alice informs Bob which photons she has selected to pass through $U^\theta$ once. This allows Bob to estimate the parameter value based on the intersection of the two sets of passes. Notably, Fisher information is additive, i.e., $F(\theta)= \sum_{i=1}^{m}F(\theta_i)$, where $F(\theta_i)$ represents the Fisher information of the $i$-th subsystem. Therefore, the Fisher information becomes $F=p\cdot 1+(1-p)N^2$, implying that the smaller $p$ is, the closer the estimation precision is to $1/N^2$. Here, we set $p=0.1$, meaning that 10\% of the photons in sequence $S_A$ pass $U^\theta$ once, while 90\% of photons pass $U^\theta$ $N$ times. So the quantum Cram\'{e}r-Rao bound approaches the Heisenberg limit, denoted as
\begin{equation} \label{eq:delta2theta}
  \delta^2\theta\rightarrow \frac{1}{p_emN^2}.
\end{equation}

To illustrate this method more intuitively, we utilize the likelihood function curve of system $\rho$. Based on the non-zero outcomes with different probabilities in Table~\ref{tab:tab1}, we derive the likelihood function of $\rho$ is
\begin{equation}
  \begin{aligned}
    L(\theta,N)\propto& (1+{\rm cos}N\theta)^{n_1}(1-{\rm cos}N\theta)^{n_2}\\
  \cdot&(1+{\rm sin}N\theta)^{n_3}(1-{\rm sin}N\theta)^{n_4},
  \end{aligned}
\end{equation}
where $n_1$, $n_2$, $n_3$, and $n_4$ represent the photon numbers corresponding to their respective probabilities. Therefore, the likelihood function is given by $L=L(\theta,1)\cdot L(\theta,4)$, where we set $N=4$.

\begin{figure}[bht]
  \centering
  \includegraphics[width=\linewidth]{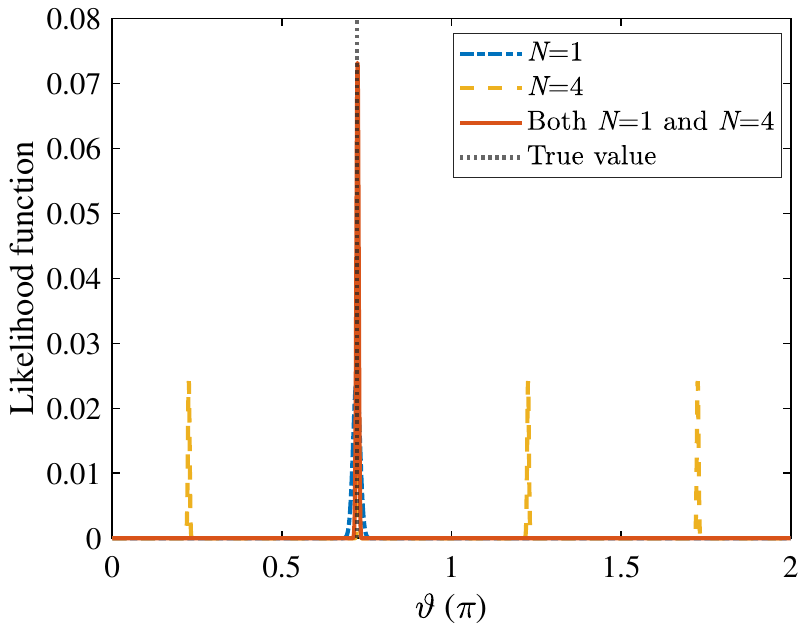}
  \caption{The likelihood functions for distinct pass strategy in the simulation of 140 EPR pairs. It is obvious that the combination of $N=1$ and $N=4$ strategy has advantages on precision and accuracy of phase estimation.}
  \label{fig4:multi-passes}
\end{figure}

In Fig.\ref{fig4:multi-passes}, we illustrate the effect of combining single-pass and multi-pass methods. We simulate and plot three likelihood function curves using the same 140 EPR pairs, each normalized. The single-pass method produces a relatively broad likelihood function around the true value of $\theta$, while the four-pass method yields narrower peaks but with a fourfold multiplicity, making it difficult to identify the true value. By combining single-pass and four-pass methods with the same quantity of qubits, we obtain a narrow peak at the correct value of $\theta$. Although the combination method results in a relatively broad peak compared to the four-pass method, it indicates a limited increase in variance. Therefore, this approach effectively ensures estimation accuracy while also approximately approaching the Heisenberg limit.

With the implementation of the refinement method, the QISAC protocol continues to demonstrate a superior sensing precision compared to other quantum remote sensing protocols. For instance, in the experiment of Ref.~\cite{yin2020experimental}, a quantum remote sensing system utilized 60000 entangled photon pairs to achieve a sensing precision of approximately 0.002 rad for Alice. While in the QISAC protocol, these same 60000 entangled photon pairs can simultaneously transmit 48000 bits of information while maintaining a sensing precision of approximately 0.0012 rad, which is a 40\% reduction compared to 0.002 rad. This precision value is calculated using Eq.~(\ref{eq:delta2theta}) with parameters set as $p_e=80\%$, $m=60000$, and $N=4$.

\section{Security analysis}
\label{sec:security}
The security of our QISAC protocol relies on two crucial assumptions. We assume that the shared Bell states are independent and identically distributed (i.i.d.), and Eve performs the same attack every time. Besides, authentication between Alice and Bob is assumed to be guaranteed. With these two assumptions in place, Alice and Bob can resist against Eve's man-in-the-middle attack. In this scenario, Eve can entangle her auxiliary system $E$ with system $B$ by applying $U_{\text{Eve}}$, after which she sends system $B$ to Bob. According to the quantum De Finetti theorem~\cite{renner2007symmetry}, the joint state of systems $B$ and $E$ asymptotically resembles a direct product of i.i.d. subsystems. Thus, it is sufficient to analyze an individual subsystem separately. To assess the security of the protocol, we divide it into two aspects: the security of information bit and the security of parameter estimation, employing Wyner's wiretap channel theory~\cite{wyner1975wire} and asymmetric Fisher information gain, respectively.

\subsection{Securiy of information bit}
The extent of eavesdropping can be assessed by examining the QBER. In this section, we aim to determine the secrecy capacity bound for our scheme. According to the Wyner's wiretap channel theory~\cite{wyner1975wire}, there exists a coding method that facilitates secure and reliable information transmission at a rate below the secrecy capacity, providing the secrecy capacity is positive. The secrecy capacity of secure communication for our scheme is 
\begin{equation}
  C_s=C_m-C_w,
\end{equation}
where $C_m$ and $C_w$ represent the channel capacities of the main and wiretap channels, respectively. To evaluate the wiretap channel capacity, we will examine the specific process of eavesdropping. Without loss of generality, we can define the attack as an identical and individual unitary transformation $U_{\text{Eve}}$ applied to the emitting state by Alice in the channel, representing a collective attack scenario. The mathematical description of an EPR pair's state after the first transmission is
\begin{equation}
  \rho_{AB}={\text{Tr}}_E(U_{\rm Eve}\lvert E \rangle \lvert \psi^- \rangle \langle \psi^- \rvert \langle E \rvert U_{\rm Eve}^\dagger).
\end{equation}
We can eliminate all the nondiagonal elements of $\rho_{AB}$ in Bell basis~\cite{kraus2005lower} based on introducing operators that two users both apply the same transformation randomly from $\left \{ I,\sigma_x,\sigma_y,\sigma_z \right \} $, and has the simple form $\rho_{AB}=\lambda_1 p_{\lvert\psi^-\rangle}+\lambda_2 p_{\lvert\psi^+\rangle}+\lambda_3 p_{\lvert\phi^-\rangle}+\lambda_4 p_{\lvert\phi^+\rangle}$, where $p_{\lvert i\rangle}=\lvert i\rangle \langle i\rvert$. The parameter $\lambda_i$ is controlled by the malicious third party, Eve.
Thus, the purification $\lvert \psi_{ABE}\rangle$ of state $\rho_{AB}$ is
\begin{equation}
  \lvert \Psi_{ABE}\rangle=\sum_{i=1}^{4} \sqrt{\lambda_i}\lvert \Psi_i \rangle \lvert E_i \rangle,
\end{equation}
where ${\lvert \Psi_i \rangle}$ is the Bell state of system $AB$ and ${\lvert E_i \rangle}$ is the set of orthogonal states of system $E$. The parameter $\lambda_i$ is determined by the QBER $\varepsilon_x$, $\varepsilon_y$ and $\varepsilon_z$ for the $\sigma_x$, $\sigma_y$ and $\sigma_z$ bases in first transmission, respectively, that is, $\varepsilon_x=\lambda_2+\lambda_4$, $\varepsilon_y=\lambda_2+\lambda_3$ and $\varepsilon_z=\lambda_3+\lambda_4$. Tracing out system $B$ from system $ABE$, we can get system $AE$ is
\begin{equation}
  \rho_{AE}={\rm Tr}_B(\lvert \Psi_{ABE}\rangle\langle \Psi_{ABE} \rvert)=\frac{1}{2}(p_{\lvert \psi\rangle}+p_{\lvert \phi\rangle}),
\end{equation}
where
\begin{equation}
  \begin{aligned}
  \lvert \psi \rangle&= \sqrt{\lambda_3}\lvert 0\rangle_A \lvert E_3 \rangle+\sqrt{\lambda_4}\lvert 0\rangle_A \lvert E_4 \rangle\\
&+\sqrt{\lambda_2}\lvert 1\rangle_A \lvert E_2 \rangle-\sqrt{\lambda_1}\lvert 1\rangle_A \lvert E_1 \rangle,\\
\lvert \phi \rangle& = \sqrt{\lambda_1}\lvert 0\rangle_A \lvert E_1 \rangle+\sqrt{\lambda_2}\lvert 0\rangle_A \lvert E_2 \rangle\\
&-\sqrt{\lambda_3}\lvert 1\rangle_A \lvert E_3 \rangle+\sqrt{\lambda_4}\lvert 1\rangle_A \lvert E_4 \rangle.
  \end{aligned}
\end{equation}
After the first transmission, Alice implies system $A$ senses the unknown system firstly and then encodes information through performing $U_0$ and $U_1$. So the operators are $U_0^{N\theta}=U_0\cdot U^{N\theta}$ and $U_1^{N\theta}=U_1\cdot U^{N\theta}$, respectively. The encoded states become
\begin{gather}
  \label{eq:rhoae1}
  \rho_{AE,0}^{N\theta}=(U_0^{N\theta}\otimes\mathbb{1})\rho_{AE}(U_0^{N\theta}\otimes\mathbb{1})^\dagger,\\
  \label{eq:rhoae2}
  \rho_{AE,1}^{N\theta}=(U_1^{N\theta}\otimes\mathbb{1})\rho_{AE}(U_1^{N\theta}\otimes\mathbb{1})^\dagger,
\end{gather} 
where $\mathbb{1}$ represents a 4-dimensional identity matrix.
If the quantum channel is modeled as a depolarizing channel $\mathcal{E}: \rho \mapsto pI/2+(1-p)\rho$, the QBERs in the first eavesdropping check are $e_x=e_y=e_z=p/2$. We assume that each qubit has the same distribution, and the partner photons transmit through a depolarizing channel. Thus, we apply the Holevo bound to obtain the mutual information between Alice and Eve. The upper bound on the information that Eve can steal is
\begin{equation}
  \begin{aligned}
  I(A:E) &\le S(\sum_{i}p_i\rho_{AE,i}^{N\theta})-\sum_{i}p_{i}S(\rho_{AE,i}^{N\theta})\\
&=h(e),
  \end{aligned}
\end{equation}
where $e$ is QBER in first transmission. Here, we denote the binary Shannon entropy by $h(x)=-x{\rm log}_2x-(1-x){\rm log}_2(1-x)$.

The capacity of the main channel $C_m$ can be described based on the two QBERs in both transmissions. We reasonably assume that our two-qubit encoding scheme is a two-fold tensor product of a single depolarizing channel. In other words, after both particles of an EPR pair pass through the channel, the probability of the depolarizing channel $\mathcal{E}^{\otimes2}$ becomes $2p-p^2$. Since Bob measures the EPR pairs in two observables separately, we can compute the mutual information between Alice and Bob in two measurement bases, which are
\begin{equation}
  I(A:B)_{1}=I(A:B)_{2}=1-h[2e(1-e)].
\end{equation}
We find that the two mutual informations performing $\hat{O_1}$ and $\hat{O_2}$ are equal. Thus, the lower bound of the secrecy capacity is 
\begin{equation}
\label{eq:CsQISAC}
  C_s\ge 1-h[2e(1-e)]-h(e).
\end{equation}

\begin{figure}[bht]
  \centering
  \includegraphics[width=\linewidth]{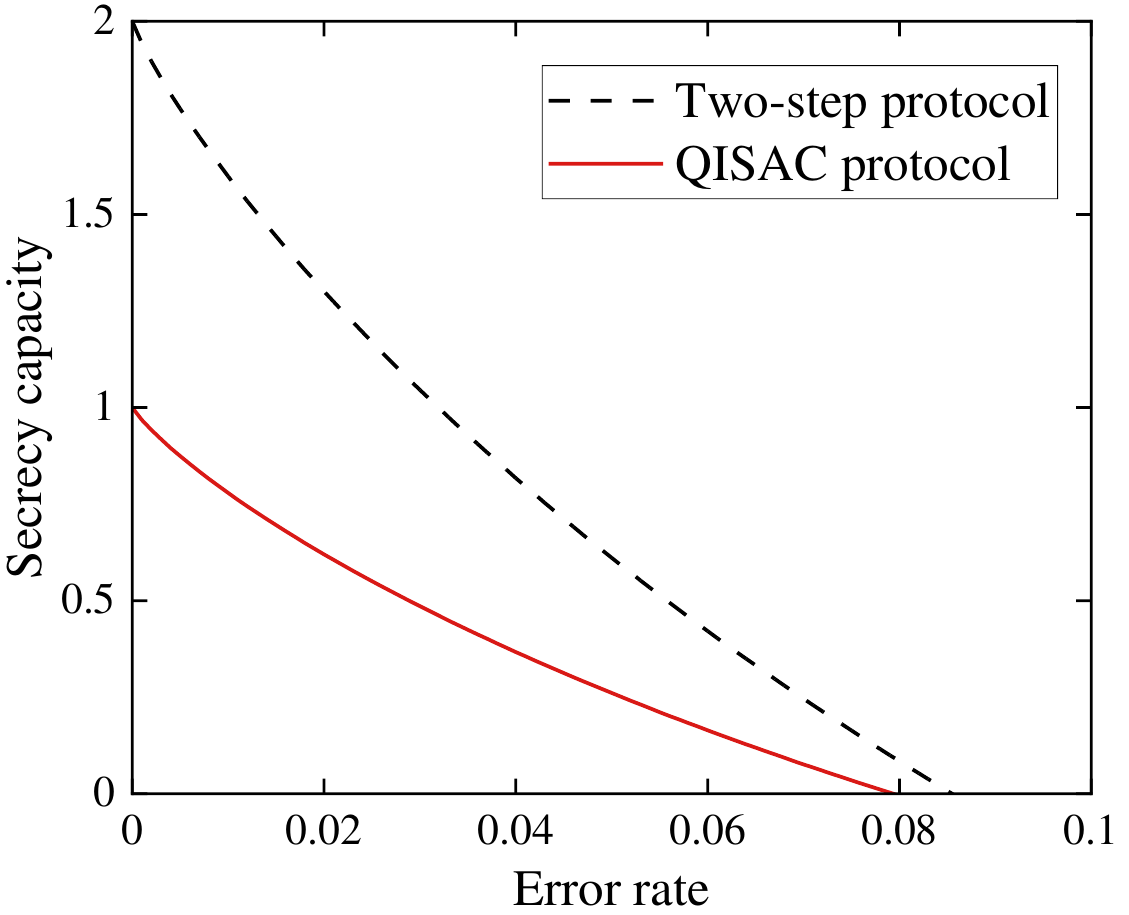}
  \caption{Comparison of the secrecy capacity between two-step QSDC protocol and our QISAC protocol. The abscissa is QBER of eavesdropping check, corresponding to the parameter $p/2$ in the depolarizing channel.} 
  \label{fig5:secrecy capacity}
\end{figure}

In the simplification scenario where $Q_B=Q_E=1$, the tight estimation of the secrecy capacity for modified two-step QSDC in the depolarizing channel is $C_s \geq 2-h_4(\textbf{e})-2h(\varepsilon)$, as shown by Eq.~(\ref{eq:twostep}). We plot and compare the secrecy capacity $C_s$ versus the QBER $e$ between our scheme and the modified two-step QSDC protocol in Fig.~\ref{fig5:secrecy capacity}. The result shows that the entire curve of our QISAC scheme lies below that of the two-step QSDC protocol. At low error rates, the latter's capacity is twice that of our scheme, but it rapidly diminishes to zero as the error rate increases. To be more specific, the QBER threshold for our QISAC scheme is 7.9\%, while the threshold of the two-step protocol is 8.6\%. The QISAC protocol maintains secure until the QBER exceeds the computed threshold. The threshold QBER, typically employed as a metric, quantifies the extent of channel noise that a protocol can tolerate. Our selection of encoding methodology, wherein a single EPR pair carries only one bit of information instead of two utilizing dense coding, inherently renders the QISAC system less robust against channel noise. Notably, the two-step QSDC protocol~\cite{deng2003two} leverages dense coding to transmit two bits of information per EPR pair, whereas QISAC, adhering to our encoding approach, restricts the transmission to a single bit per pair. A comparative analysis of Eq.~(\ref{eq:twostep}) and Eq.~(\ref{eq:CsQISAC}) underscores this disparity, elucidating how the distinct encoding strategies lead to a lower threshold QBER for QISAC. Nevertheless, we intentionally compromise on certain aspects of secure communication performance inherent in entanglement-based QSDC protocols to incorporate quantum sensing capabilities. This integration leverages the EPR pairs to realize a converged platform for quantum communication and sensing.

Note that the double CNOT attack~\cite{fahmi2008comment} enables the distinction between $\lvert \psi^\pm \rangle$ and $\lvert \phi^\pm \rangle$ in the transmission of confidential information via Bell states. Next, we will demonstrate the robustness of the proposed QISAC protocol against such attacks. In the first transmission, Alice generates EPR pairs in $\lvert \psi^-\rangle$ and sends $S_B$ to Bob. When the particles in $S_B$ pass through the quantum channel, Eve applies the first CNOT gate on the qubit as the control bit, and her own qubit as the target qubit. After that, the system $ABE$ becomes
\begin{equation}
 \lvert\psi^-\rangle_{AB}\lvert0\rangle_{E} \stackrel{\rm CNOT}{\longrightarrow} \frac{1}{\sqrt{2}}(\lvert01\rangle \lvert1\rangle-\lvert10\rangle \lvert0\rangle)_{ABE}.
\end{equation}
Tracing out system $E$ from $ABE$, we can get system $AB$ is
\begin{equation}
  \rho_{AB}={\rm Tr}_E(\rho_{ABE})=\frac{1}{2}(\lvert01\rangle\langle01\rvert+\lvert10\rangle\langle10\rvert).
\end{equation}
If Alice and Bob both measure in $\sigma_x$ basis, the probabilities $P(A,B)$ of outcomes $+$ or $-$ are
\begin{equation}
  P(+,+)=P(+,-)=P(-,+)=P(-,-)=\frac{1}{4}.
\end{equation}
Without the double CNOT attack from Eve, Alice and Bob get deterministic opposite results,
\begin{equation}
  P(+,-)=P(-,+)=\frac{1}{2}.
\end{equation}
In other word, the double CNOT attack would result in a 50\% QBER when measuring in the $\sigma_x$-basis. Similarly, when both Alice and Bob measure in the $\sigma_y$-basis, they also encounter a 50\% QBER. However, measuring in the $\sigma_z$-basis does not introduce errors during the eavesdropping check. 

In the depolarizing channel, the QBERs $\varepsilon_x$, $\varepsilon_y$ and $\varepsilon_z$ should be approximately equal. Consequently, once two legitimate users observe that $\varepsilon_x$ and $\varepsilon_y$ deviate significantly from $\varepsilon_z$, they will terminate the transmission and revert to step 1. This approach effectively counters the double CNOT attack launched by malicious eavesdroppers.

\subsection{Security of phase estimation}
As demonstrated in the quantum remote sensing protocol~\cite{yin2020experimental}, the security level of phase estimation is characterized by the asymmetric Fisher information gain. To assess how much Fisher information Bob acquires in the protocol under depolarizing circumstances, we calculate a new set of probabilities for two observables, where the non-zero value $1\pm\cos N\theta$ becomes $1\pm(1-2e)^2\cos N\theta$ and $1\pm\sin N\theta$ becomes $1\pm(1-2e)^2\sin N\theta$, the other probabilities are still $0$. Substituting the probabilities $P(x_i|\theta)$ of Eq.~(\ref{eq:cfi}) with new data, we obtain
\begin{gather}
  \label{eq:noisy-cfi1}
F^{\hat{O_1}}=\frac{N^2(1-2e)^4{\rm sin}^2N\theta}{2[1-(1-2e)^4{\rm cos}^2\theta]},\\
  \label{eq:noisy-cfi2}
F^{\hat{O_2}}=\frac{N^2(1-2e)^4{\rm cos}^2N\theta}{2[1-(1-2e)^4{\rm sin}^2\theta]}.
\end{gather}
Clearly, the level of Fisher information can be assessed via the QBER. Since Fisher information is additive, the Fisher information Bob gains satisfies
\begin{widetext} 
\begin{align}
  F_{\text{Bob}}=F^{\hat{O_1}}+F^{\hat{O_2}}=\frac{N^2(1-2e)^4}{2}\frac{1-(1-2e)^4+2(1-2e)^4{\rm cos}^2\theta {\rm sin}^2\theta}{1-(1-2e)^4+(1-2e)^8{\rm sin}^2\theta {\rm cos}^2\theta}\ge \frac{N^2(1-2e)^4}{2}.
\end{align}
\end{widetext}

\begin{figure}[bht]
  \centering
  \includegraphics[width=\linewidth]{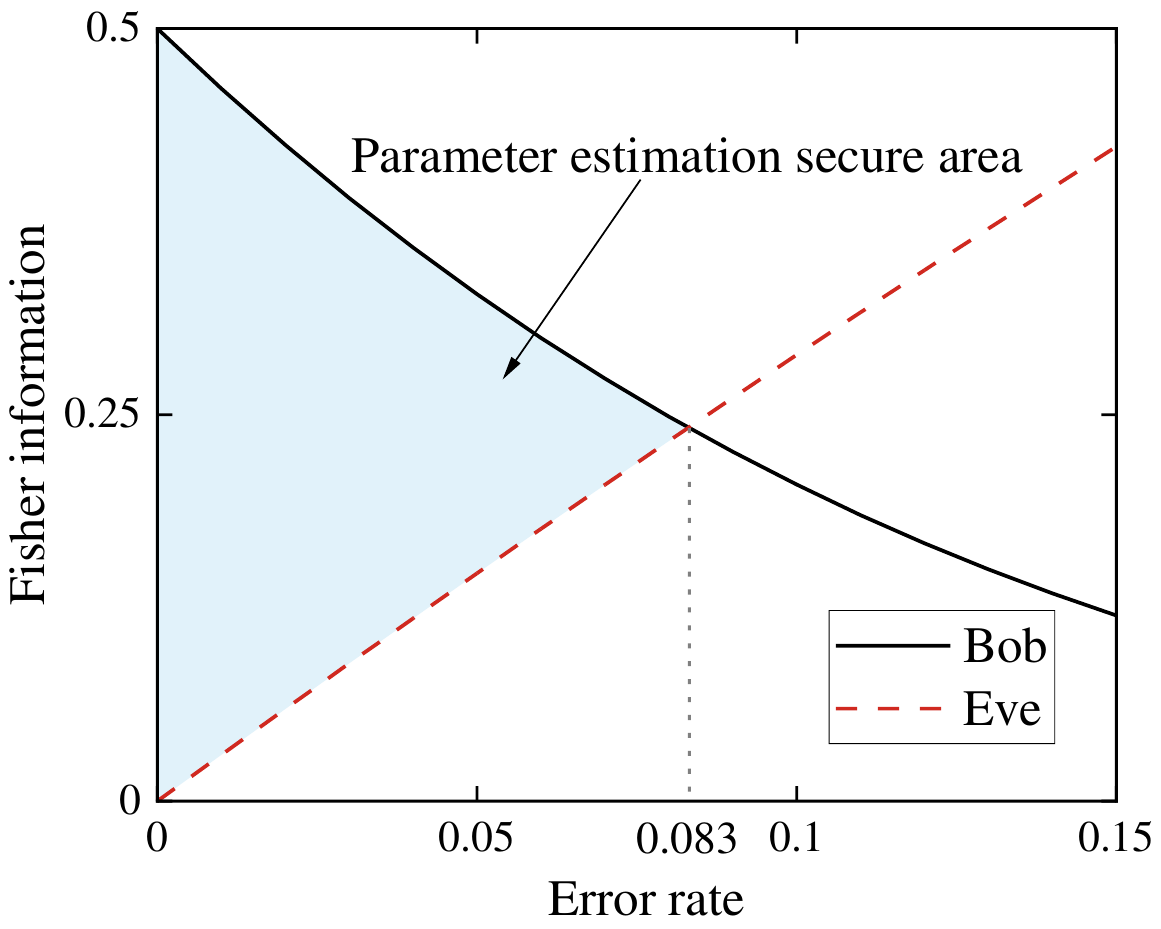}
  \caption{Fisher information of which Bob and Eve gain. The blue area between two curves is the security area for parameter estimation, and the horizontal coordinate of the intersection point is QBER threshold 8.3\%. For simplification, we set $N=1$, while QBER threshold is independent of $N$.} 
  \label{fig6:fisher}
\end{figure}

To quantify the Fisher information acquired by the eavesdropper, we assume that Eve possesses formidable capabilities, constrained only by the principles of quantum physics. Since all information available to Eve originates from Alice, we substitute the mixed state density matrix $\rho_{AE}=1/2\rho_{AE,0}^{N\theta}+1/2\rho_{AE,1}^{N\theta}$ from Eq.~(\ref{eq:rhoae1}) and Eq.~(\ref{eq:rhoae2}) into Eq.~(\ref{eq:qfi}) to compute Eve's quantum Fisher information, yielding
\begin{equation}
  F_{\text{Eve}}=\frac{N^2(3e-4e^2)}{1-e}.
\end{equation}
The Fisher information $F_{\text{Bob}}$ and $F_{\text{Eve}}$ versus the QBER are depicted in Fig.~\ref{fig6:fisher}. The blue area between these two curves constitutes an asymmetric gain, that is, if the QBER falls within this area for our integrated scheme, we can reliably guarantee the security of parameter estimation, and the QBER threshold corresponding to the intersection is 8.3\%.

Combined the two thresholds, the security of this proposal is confirmed through a purely mathematical derivation of both quantum communication and quantum remote sensing aspects. As long as the QBER during transmission remains below 7.9\%, our QISAC scheme remains both aspects secure while achieving integrated remote sensing and communication.

\section{Performance analysis} \label{sec:performance}
\subsection{Advantages of using two observables}
Let us explore the advantages of performing two observables $\hat{O}_1$ and $\hat{O}_2$. Consider an experiment with $\nu$ measurements, where the number of results for each non-zero outcome with probabilities given in Table~\ref{tab:tab1} is denoted as ${n_i}$, where $i\in \left \{ 1,2,3,4 \right \} $ and the sum of $n_i$ is equal to $\nu$. For ease of calculation, the value of $N$ in this section is 1 unless otherwise specified. Utilizing Eq.~(\ref{eq:likelihood}), which provides Bob's likelihood function for $\theta$, we arrive at 
\begin{equation}
  \begin{aligned}
    L(\theta)\propto& (1+{\rm cos}\theta)^{n_1}(1-{\rm cos}\theta)^{n_2}\\
    \cdot&(1+{\rm sin}\theta)^{n_3}(1-{\rm sin}\theta)^{n_4}.
  \end{aligned}
\end{equation}
The likelihood function of $\hat{O_1}$ is $L_1(\theta)\propto (1+\text{cos}\theta)^{n_1}(1-\text{cos}\theta)^{n_2}$ and that of $\hat{O_2}$ is $L_2(\theta)\propto (1+\text{sin}\theta)^{n_3}(1-\text{sin}\theta)^{n_4}$, while $L(\theta)$ is the product of $L_1(\theta)$ and $L_2(\theta)$.
Under the condition of utilizing 500 EPR pairs as probe states, we simulate the measurement process and plot the likelihood function curves for three different observable schemes, comparing them with the true value 0.8$\pi$.
\begin{figure}[bht]
  \centering
  \includegraphics[width=\linewidth]{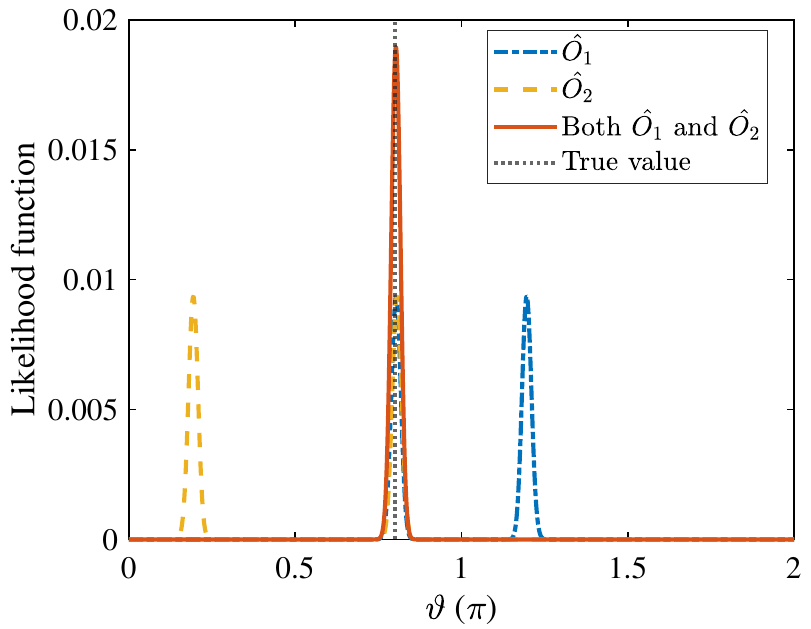}
  \caption{The normalized likelihood functions for using distinct measurement schemes. Note that the number $N$ of $U^\theta$ is set to 1.} 
  \label{fig7:likelihood}
\end{figure}

Table.~\ref{tab:tab1} reveals that the probabilities of non-zero measurement outcomes for $\hat{O_1}=\sigma_x\otimes \sigma_x$ and $\hat{O_2}=\sigma_y\otimes \sigma_x$ depend solely on $\text{cos}\theta$ and $\text{sin}\theta$, respectively. Consequently, the estimation based on either observable alone can only cover the parameter range $[0,\pi)$ due to their symmetric likelihoods within the $[0,2\pi)$ interval, as depicted in Fig.~\ref{fig7:likelihood}. This ambiguity is resolved by utilizing two sets of measurement bases orthogonal to each other on the Bloch sphere, allowing for estimation across the entire $[0,2\pi]$ range.

\begin{figure}[bht]
  \centering
  \includegraphics[width=\linewidth]{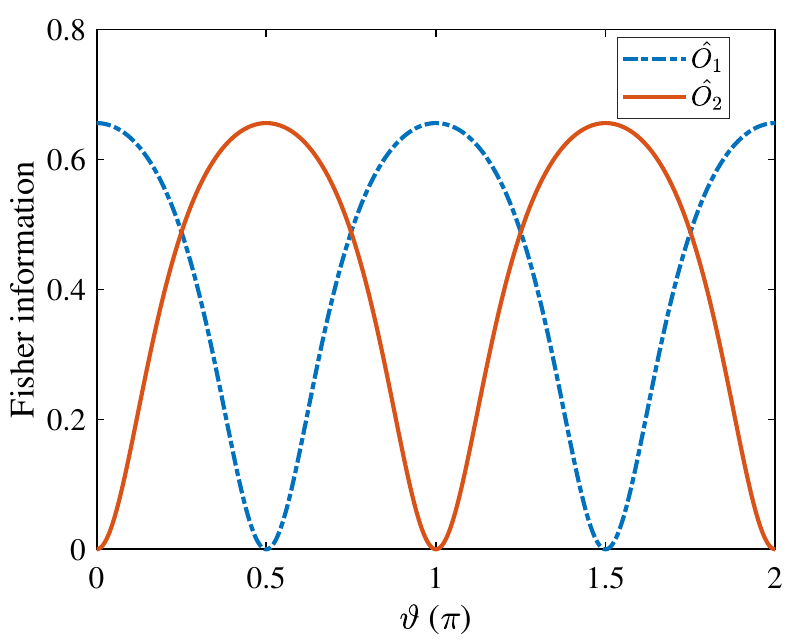}
  \caption{The classical Fisher information using $\hat{O_1}$ and $\hat{O_2}$ observables. In the simulation, $N=1$.} 
  \label{fig8:cfi-noisy}
\end{figure}

An additional reason for employing both observables lies in maintaining security within noisy quantum channel. While we have previously demonstrated that $\hat{O_1}$ and $\hat{O_2}$ are both optimal observables in ideal scenario in Sec.~\ref{sec:security}. But their optimality no longer holds in a noisy environment, and the CFIs have been previously derived and are given by Eq.~(\ref{eq:noisy-cfi1}) and Eq.~(\ref{eq:noisy-cfi2}). Obviously, these CFIs are functions of the unknown parameter $\theta$, which fluctuates in value as $\theta$ changes. With a fixed $e$ value of 5\%, we plot the change in CFI with respect to the parameter $\theta$ to be estimated in Fig.~\ref{fig8:cfi-noisy}. Although implementing two observable variables is more complex than with one, the overall benefits significantly outweigh the drawbacks.

\subsection{Limited resources}\label{sec:limited}
The available entanglement resources are limited in real-life implementation. We must consider the performance of the QISAC protocol in parameter estimation under the constraint of limited resources. In the asymptotic regime, using expected values to estimate the parameter $\theta$ yields an estimate very close to the true value, with minimal fluctuations introduced solely by inherent instrument errors. As depicted in Fig.~\ref{fig7:likelihood}, the likelihood function becomes an increasingly narrow normal distribution centered around the true value of $\theta$. However, when dealing with limited data, we need to account for potential bias in the estimate of $\theta$, which arises due to the constraints imposed by finite data availability.

\begin{figure}[bht]
  \centering
  \includegraphics[width=\linewidth]{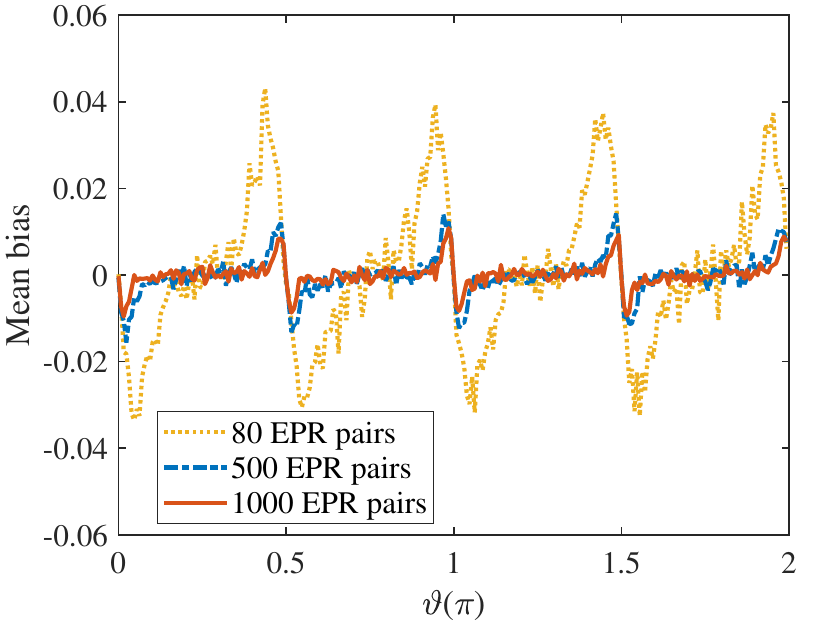}
  \caption{The mean bias between phase estimate and true value in the case of different numbers of EPR pairs. With the increase of EPR pairs number, the bias value decreases. Here we set $N=1$.}
  \label{fig9:bias}
\end{figure}

We simulate and plot the mean bias for different $\theta$ values with varying photon numbers in Fig.~\ref{fig9:bias}, where $\theta$ ranges from $0$ to $2\pi$. To mitigate systematic errors, we repeat each parameter estimation $10^3$ times, average the results, and compare them with the true value. Fig.~\ref{fig9:bias} reveals that for most $\theta$ values, the bias remains smaller than the standard deviation $1/\sqrt{\nu}$ ($N=1$), where $\nu$ represents the number of measurements, corresponding to the number of EPR pairs in the legend. This observation demonstrates the robustness of the QISAC protocol's estimation method even under finite entanglement resources, rendering it a reliable approach. Hence, the QISAC protocol can also ensure sensing accuracy while performing both sensing and communication tasks. While the improvement in accuracy does not increase linearly with the augmentation of entanglement resources. Initially, substantial gains are achieved, but as the entanglement resources expand, the accuracy  improvement becomes progressively gradual. This behavior arises from inherent limitations in the method itself.

When the true value of $\theta$ aligns with an integer or half-integer multiple of $\pi$, the estimated value of $\theta$ exhibits more significant bias, and this phenomenon becomes pronounced with fewer entangled photons. To optimize their estimation method, Alice and Bob can avoid this region by establishing an extra angle $\delta$ and shifting $\theta$ away from the problematic range.

Under the finite resource assumption, an intrinsic trade-off exists between achievable precision and security against the double CNOT attack and general man-in-the-middle attack. In the first checking round, each qubit intercepted and subjected to the CNOT gate by Eve randomly has a $1/2$ probability of introducing an error signal for Bob, provided that Bob measures the sampling qubits in either $\sigma_x$ or $\sigma_y$ basis. As previously mentioned, the legitimate parties can abandon the protocol if they detect Eve during the security check. The probability of selecting photons as eavesdropping-check samples is $(1-p_e)/2$. Assume Eve intercepts and measures every qubit to obtain information, the probability of detecting Eve against the double CNOT attack is denoted by
\begin{equation}
  P_{\text{det1}}=1-(\frac{5+p_e}{6})^{(1-p_e)m/2},
\end{equation}
where $m$ is the total number of qubits generated and distributed by Alice. In the case of the general man-in-the-middle attack, Eve intercepts and measures each qubit in $\sigma_x$, $\sigma_y$, or $\sigma_z$ basis randomly, resulting in a $1/3$ error probability respectively for Bob. If the number of qubits intercepted by Eve is $k$, the probability of detecting Eve against a general man-in-the-middle attack is
\begin{equation}
  P_{\text{det2}}=1-(\frac{5+p_e}{6})^k.
\end{equation}

\begin{figure}[bht]
  \centering
  \includegraphics[width=\linewidth]{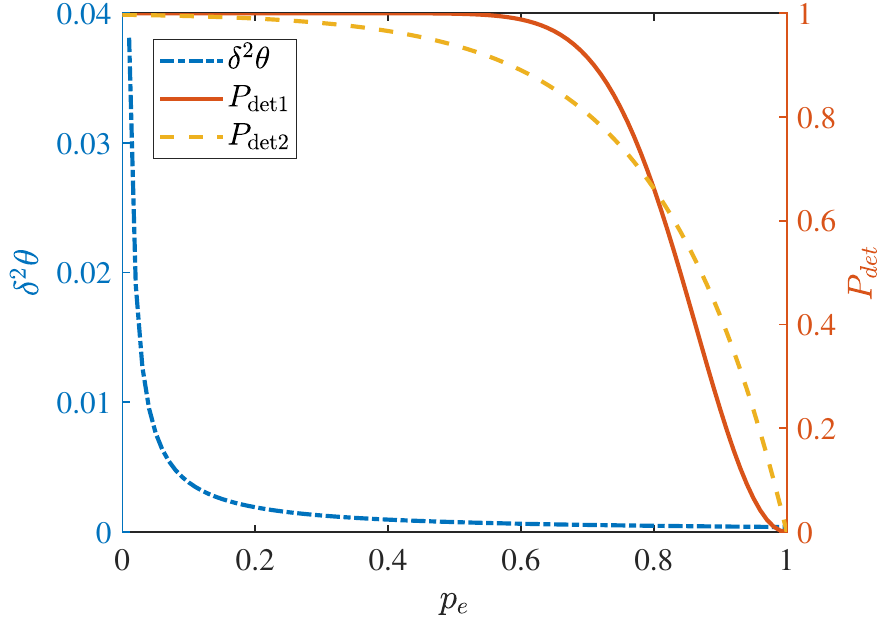}
  \caption{The precision-security trade-off curves against the double CNOT attack and general man-in-the-middle attack, where the blue dot dash curve represents variance of $\theta$, the red solid curve and the yellow dotted curve represent the probability of detecting Eve against the double CNOT attack and general man-in-the-middle attack, respectively. Note that we set the total EPR pairs number $m$ as 320 and qubits intercepted by Eve $k$ equals 32. The $N$ of multi-round strategy is 3.} 
  \label{fig10:tradeoff}
\end{figure}

Fig.~\ref{fig10:tradeoff} provides a detailed illustration of the precision-security trade-off relationship. As the encoded ratio $p_e$ increases, more resources are allocated to the parameter estimation process, resulting in reduced variance of the estimate. However, this allocation comes at the cost of diminished security, as fewer resources remain for security checking. When the value of $p_e$ is small, the solid curve $P_{\text{det1}}$ representing resistance to the double CNOT attack exhibits a gradual descent rate, and $P_{\text{det1}}$ remains around $1$ until the $p_e$ value reaches $0.6$. At this point, the QISAC protocol demonstrates robust resistance to the double CNOT attack. Interestingly, as the $p_e$ value increases further, the $P_{\text{det1}}$ curve experiences a sudden drop and intersects with $P_{\rm det2}$, a dashed curve that provides defense against the general man-in-the-middle attack. When $p_e$ becomes large, the QISAC protocol exhibits better resistance to a general man-in-the-middle attack than to a double CNOT attack. Conventionally, we can optimize one aspect while considering the impact on the other, achieve a desired trade-off between security and precision.

\subsection{Optimal Value of $N$}
\begin{figure*}[htb]
  \centering
  \includegraphics[width=\linewidth]{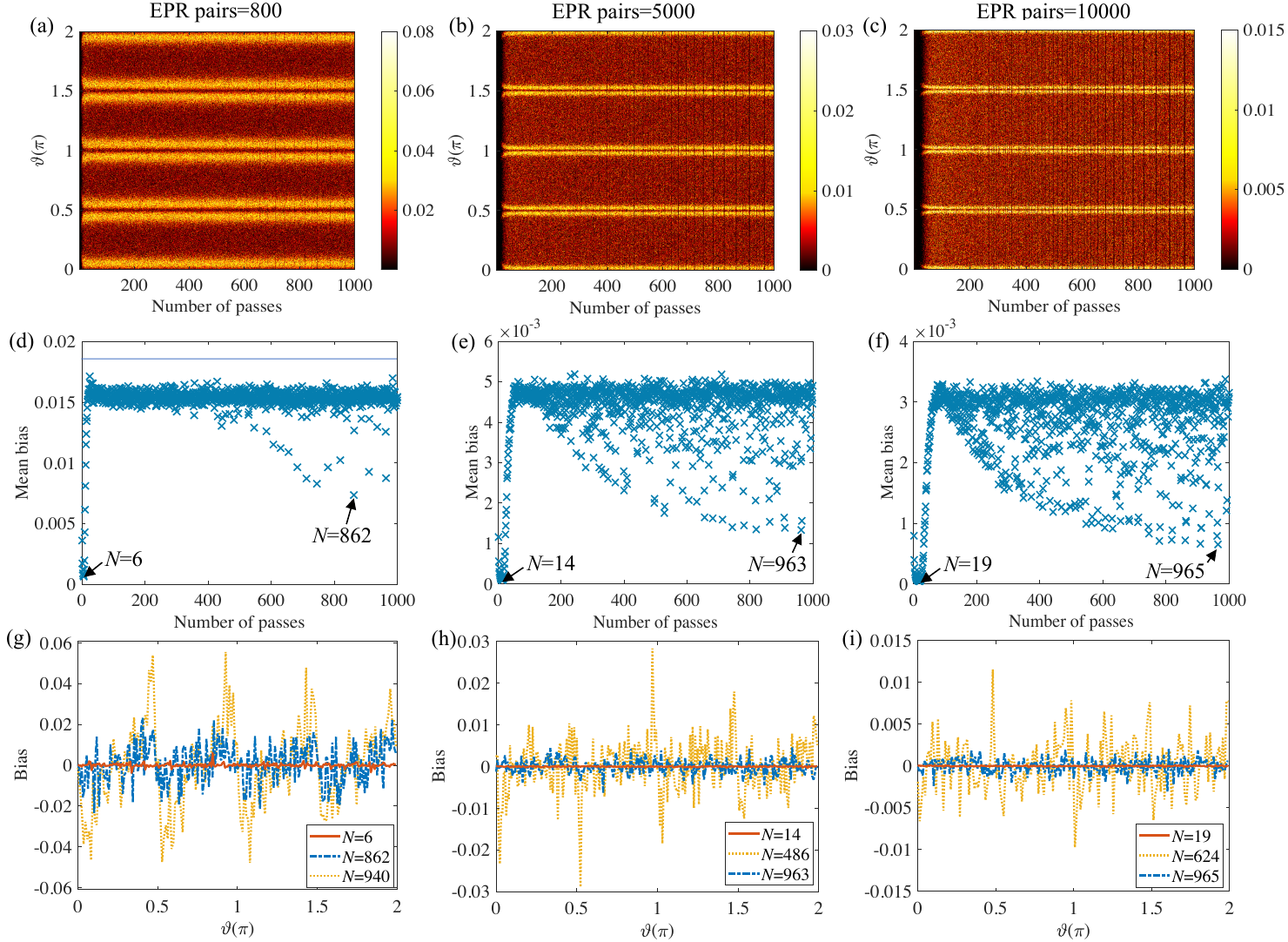}
  \caption{Simulation results to determine the optimal number of passes. (a), (b), and (c) are heatmaps of estimation bias, each with 800, 5000 and 10000 EPR pairs, respectively. Each combination of the number of passes and the true value of  $\theta$ is input into the estimation model to obtain the estimated bias value, which is then converted into  a color block according to the colorbar. (d), (e), and (f) are the auxiliary scatter diagrams, where each point represents the mean bias averaged over the parameter's true value from 0 to 2$\pi$, with different numbers of passes. (g), (h), and (i) are bias curves for specific $N$ values. We have chosen these quantities of EPR pairs, because their 10\% fraction of photons ($N$=1) matches the quantity of EPR pairs used in Fig.~\ref{fig9:bias}, where the mean bias curve is obtained with all EPR pairs sensing only once ($N=1$), which facilitates comparisons between them.}
  \label{fig11:N}
\end{figure*}

In Sec.~\ref{sec:protocol}, we employ a multi-round strategy to saturate the Heisenberg limit and refine the proposed QISAC protocol which allocates 10\% of photons in sequence $S_A$ to pass through $U^\theta$ only once. However, from the perspective of estimation accuracy, a larger number of passes $N$ in the multi-round strategy does not necessarily yield better results. In Fig.~\ref{fig11:N}, we simulate the variation of bias with the number of passes for three different quantities of EPR pairs: 800, 5000, and 10000 pairs. The simulation results include heatmaps showing bias variation with $N$ and the true value of $\theta$, scatter plots of the mean bias averaged over true values of $\theta$ from 0 to 2$\pi$, and curves depicting how bias changes with different true values of $\theta$.

Fig.~\ref{fig11:N}.~(a), (b), and (c) are heatmaps of estimation bias for 800, 5000 and 10000 EPR pairs, respectively. Every color block represents a certain bias value corresponding to the colorbar, obtained by inputting the $N$ value and $\theta$ value into the estimation for simulation. As shown in Fig.~\ref{fig11:N}.~(a), (b), and (c), the bias remains minimal when $N$ is small ($N<30$) and gradually increases with the increase of $N$. However, all three heatmaps exhibit some black vertical stripes near the end of the x-axis. As the number of EPR pairs increases, these black stripes become more numerous and more pronounced. These black stripes indicate that the estimation bias becomes smaller for such values of $N$, regardless of the true value of $\theta$. This suggests that these $N$ values might be optimal, which allows the model to maintain a low bias and high accuracy. In contrast, other $N$ values show significant bias fluctuations near half-integer multiples of $\pi$, which are represented by yellow horizontal stripes in the heatmaps.

To more clearly determine which $N$ is the optimal value, we present the auxiliary scatter diagrams in Fig.~\ref{fig11:N}.~(d), (e), and (f). Each point in these diagrams represents the mean bias averaged over $\theta$ from 0 to 2$\pi$ with different numbers of passes. In Fig.~\ref{fig11:N}.~(d), (e), and (f), we mark the $N$ values corresponding to the smallest and the second-smallest mean bias, respectively. Obviously, with the increase in the number of EPR pairs, the average deviation value decreases as a whole, and the $N$ value corresponding to the minimum average deviation increases. Furthermore, the bias trends in the scatter plots correspond to the heatmaps. It is because after multiple sensing of the system, the likelihood function produces several narrow peaks between 0 and 2$\pi$. When these are multiplied by the peak at $N=1$, the estimation becomes more convergent, as shown in Fig.~\ref{fig4:multi-passes}. However, as the value of $N$ continues to increase, the peaks become very dense between 0 and 2$\pi$, making the estimation accuracy almost entirely dependent on N=1, \textit{i.e.} the 10\% of EPR pairs used for estimation. Therefore, the final bias value tends to stabilize. Our calculations show that the mean bias values in Fig.~\ref{fig9:bias} are consistent with the stable mean bias values in Fig.~\ref{fig11:N}.~(d), (e), and (f), which confirms our analysis. The sudden decrease in mean bias at certain $N$ values is intriguing. We tentatively hypothesize that it is related to the finite sample size, enhancing the estimation accuracy for certain $N>1$ estimation processes.

It can lead us to identify a sufficiently large optimal value of $N$, large enough to demonstrate the quantum advantage in variance~\cite{zheng2019ab}. To verify the accuracy of the estimation for the large optimal value of $N$, we have drawn bias curves in Fig.~\ref{fig11:N}.~(g), (h), and (i). These figures show the bias curves obtained from the estimation models, with $N$ values related to the smallest, second smallest, and stabilizing mean bias, respectively, for different numbers of EPR pairs. Taking Fig.~\ref{fig11:N}.~(g) for 800 EPR pairs as an example, the bias curve for $N=6$ has the smallest fluctuation, with a maximum bias of approximately 0.008. The bias curve for $N=862$ has a relatively small fluctuation, performing well despite not being as smooth as the curve for $N=6$, with a maximum bias of only 0.02, about 36\% of the maximum bias for $N=940$, where the mean bias stabilizes. This indicates that the estimation model with $N=862$ ensures a certain degree of accuracy while demonstrating quantum advantage in variance, approaching the Heisenberg limit, which is our desired large optimal $N$ value. For 5000 and 10000 EPR pairs, we found the optimal $N$ values to be $N=963$ and $N=965$, respectively. They meet both the criteria of having a quantum advantage and providing sufficiently high estimation accuracy.

Furthermore, as the number of EPR pairs increases, the bias with the optimal value of $N$ becomes less regular and more random. This suggests that, in practical applications, it depends on the number of EPR pairs whether it is necessary to allocate a small portion of photons to first roughly estimate the range of the true value and apply additional shift angle $\delta$ mentioned in Sec.~\ref{sec:limited}.

\section{Conclusions and outlook}
\label{sec:conclusion}
We propose a QISAC protocol, which enables distant communication parties to achieve secure and reliable quantum sensing alongside quantum-secured direct communication. In traditional approaches, two remote parties would require separate sets of entangled photons for sensing and communication. However, the QISAC protocol integrates two functions, utilizing photons that would otherwise be allocated to a single task. By transmitting EPR pairs, the precision of QISAC's remote sensing can approach the Heisenberg limit. In the presence of a depolarizing noisy quantum channel, we demonstrate that the secrecy capacity and asymmetric information gain ensure the security of the QISAC protocol from both informational transmission and sensing-parameter perspectives. Through the performance analysis model we provide, various adjustment strategies are proposed to optimize the performance of the protocol. The simulation of sensing performance shows that the QISAC protocol maintains superior sensing capability even while performing communication.

In the QISAC protocol, while assuring end-node security similar to other point-to-point communication protocols, it offers a distinct advantage over scenarios where one party conducts local quantum sensing followed by quantum key distribution to transmit the parameter estimate. QISAC accomplishes both tasks in a single transmission and entangled state measurement, whereas the latter approach involves three distinct steps: (1) obtaining data through local sensing; (2) using quantum key distribution to negotiate a secret key; (3) using the secret key to encrypt the sensing data and transmitting the encrypted data via classical communication. The latter approach is more cumbersome than the QISAC protocol, therefore requiring more quantum state resources as well as classical and quantum communication resources. In contrast, QISAC achieves sensing and communication more efficiently by leveraging the entanglement properties of EPR pairs to maximize resource utilization.

An intriguing direction is to experimentally demonstrate this protocol and gradually enhance its capabilities. In the future, the quantum internet will be established gradually~\cite{wehner2018quantum}. By considering both communicating parties in the protocol as two nodes in the network, QISAC would further enrich the functionality of quantum Internet.

\section{Acknowledgements}
We acknowledge support by the National Natural Science Foundation of China (Grants No. 12205011 and 62131002), the Open Research Fund Program of the State Key Laboratory of Low-Dimensional Quantum Physics under Grant No. KF202205, Beijing Advanced Innovation Center for Future Chip (ICFC). We acknowledge the helpful discussions with Dr. Sean William Moore.

\normalem

\begin{thebibliography}{0}%
\makeatletter
\providecommand \@ifxundefined [1]{%
 \@ifx{#1\undefined}
}%
\providecommand \@ifnum [1]{%
 \ifnum #1\expandafter \@firstoftwo
 \else \expandafter \@secondoftwo
 \fi
}%
\providecommand \@ifx [1]{%
 \ifx #1\expandafter \@firstoftwo
 \else \expandafter \@secondoftwo
 \fi
}%
\providecommand \natexlab [1]{#1}%
\providecommand \enquote  [1]{``#1''}%
\providecommand \bibnamefont  [1]{#1}%
\providecommand \bibfnamefont [1]{#1}%
\providecommand \citenamefont [1]{#1}%
\providecommand \href@noop [0]{\@secondoftwo}%
\providecommand \href [0]{\begingroup \@sanitize@url \@href}%
\providecommand \@href[1]{\@@startlink{#1}\@@href}%
\providecommand \@@href[1]{\endgroup#1\@@endlink}%
\providecommand \@sanitize@url [0]{\catcode `\\12\catcode `\$12\catcode
  `\&12\catcode `\#12\catcode `\^12\catcode `\_12\catcode `\%12\relax}%
\providecommand \@@startlink[1]{}%
\providecommand \@@endlink[0]{}%
\providecommand \url  [0]{\begingroup\@sanitize@url \@url }%
\providecommand \@url [1]{\endgroup\@href {#1}{\urlprefix }}%
\providecommand \urlprefix  [0]{URL }%
\providecommand \Eprint [0]{\href }%
\providecommand \doibase [0]{https://doi.org/}%
\providecommand \selectlanguage [0]{\@gobble}%
\providecommand \bibinfo  [0]{\@secondoftwo}%
\providecommand \bibfield  [0]{\@secondoftwo}%
\providecommand \translation [1]{[#1]}%
\providecommand \BibitemOpen [0]{}%
\providecommand \bibitemStop [0]{}%
\providecommand \bibitemNoStop [0]{.\EOS\space}%
\providecommand \EOS [0]{\spacefactor3000\relax}%
\providecommand \BibitemShut  [1]{\csname bibitem#1\endcsname}%
\let\auto@bib@innerbib\@empty
\end{thebibliography}%


\begin{thebibliography}{45}%
  \makeatletter
  \providecommand \@ifxundefined [1]{%
   \@ifx{#1\undefined}
  }%
  \providecommand \@ifnum [1]{%
   \ifnum #1\expandafter \@firstoftwo
   \else \expandafter \@secondoftwo
   \fi
  }%
  \providecommand \@ifx [1]{%
   \ifx #1\expandafter \@firstoftwo
   \else \expandafter \@secondoftwo
   \fi
  }%
  \providecommand \natexlab [1]{#1}%
  \providecommand \enquote  [1]{``#1''}%
  \providecommand \bibnamefont  [1]{#1}%
  \providecommand \bibfnamefont [1]{#1}%
  \providecommand \citenamefont [1]{#1}%
  \providecommand \href@noop [0]{\@secondoftwo}%
  \providecommand \href [0]{\begingroup \@sanitize@url \@href}%
  \providecommand \@href[1]{\@@startlink{#1}\@@href}%
  \providecommand \@@href[1]{\endgroup#1\@@endlink}%
  \providecommand \@sanitize@url [0]{\catcode `\\12\catcode `\$12\catcode
    `\&12\catcode `\#12\catcode `\^12\catcode `\_12\catcode `\%12\relax}%
  \providecommand \@@startlink[1]{}%
  \providecommand \@@endlink[0]{}%
  \providecommand \url  [0]{\begingroup\@sanitize@url \@url }%
  \providecommand \@url [1]{\endgroup\@href {#1}{\urlprefix }}%
  \providecommand \urlprefix  [0]{URL }%
  \providecommand \Eprint [0]{\href }%
  \providecommand \doibase [0]{https://doi.org/}%
  \providecommand \selectlanguage [0]{\@gobble}%
  \providecommand \bibinfo  [0]{\@secondoftwo}%
  \providecommand \bibfield  [0]{\@secondoftwo}%
  \providecommand \translation [1]{[#1]}%
  \providecommand \BibitemOpen [0]{}%
  \providecommand \bibitemStop [0]{}%
  \providecommand \bibitemNoStop [0]{.\EOS\space}%
  \providecommand \EOS [0]{\spacefactor3000\relax}%
  \providecommand \BibitemShut  [1]{\csname bibitem#1\endcsname}%
  \let\auto@bib@innerbib\@empty
  \bibitem [{\citenamefont {Horodecki}\ \emph {et~al.}(2009)\citenamefont
    {Horodecki}, \citenamefont {Horodecki}, \citenamefont {Horodecki},\ and\
    \citenamefont {Horodecki}}]{horodecki2009quantum}%
    \BibitemOpen
    \bibfield  {author} {\bibinfo {author} {\bibfnamefont {R.}~\bibnamefont
    {Horodecki}}, \bibinfo {author} {\bibfnamefont {P.}~\bibnamefont
    {Horodecki}}, \bibinfo {author} {\bibfnamefont {M.}~\bibnamefont
    {Horodecki}},\ and\ \bibinfo {author} {\bibfnamefont {K.}~\bibnamefont
    {Horodecki}},\ }\bibfield  {title} {\bibinfo {title} {Quantum entanglement},\
    }\href {https://doi.org/10.1103/RevModPhys.81.865} {\bibfield  {journal}
    {\bibinfo  {journal} {Reviews of Modern Physics}\ }\textbf {\bibinfo {volume}
    {81}},\ \bibinfo {pages} {865} (\bibinfo {year} {2009})}\BibitemShut
    {NoStop}%
  \bibitem [{\citenamefont {Ekert}(1991)}]{ekert1991quantum}%
    \BibitemOpen
    \bibfield  {author} {\bibinfo {author} {\bibfnamefont {A.~K.}\ \bibnamefont
    {Ekert}},\ }\bibfield  {title} {\bibinfo {title} {Quantum cryptography based
    on {B}ell’s theorem},\ }\href {https://doi.org/10.1103/PhysRevLett.67.661}
    {\bibfield  {journal} {\bibinfo  {journal} {Physical Review Letters}\
    }\textbf {\bibinfo {volume} {67}},\ \bibinfo {pages} {661} (\bibinfo {year}
    {1991})}\BibitemShut {NoStop}%
  \bibitem [{\citenamefont {Hillery}\ \emph {et~al.}(1999)\citenamefont
    {Hillery}, \citenamefont {Bu{\v{z}}ek},\ and\ \citenamefont
    {Berthiaume}}]{hillery1999quantum}%
    \BibitemOpen
    \bibfield  {author} {\bibinfo {author} {\bibfnamefont {M.}~\bibnamefont
    {Hillery}}, \bibinfo {author} {\bibfnamefont {V.}~\bibnamefont
    {Bu{\v{z}}ek}},\ and\ \bibinfo {author} {\bibfnamefont {A.}~\bibnamefont
    {Berthiaume}},\ }\bibfield  {title} {\bibinfo {title} {Quantum secret
    sharing},\ }\href {https://doi.org/10.1103/PhysRevA.59.1829} {\bibfield
    {journal} {\bibinfo  {journal} {Physical Review A}\ }\textbf {\bibinfo
    {volume} {59}},\ \bibinfo {pages} {1829} (\bibinfo {year}
    {1999})}\BibitemShut {NoStop}%
  \bibitem [{\citenamefont {Xiao}\ \emph {et~al.}(2004)\citenamefont {Xiao},
    \citenamefont {Long}, \citenamefont {Deng},\ and\ \citenamefont
    {Pan}}]{xiao2004efficient}%
    \BibitemOpen
    \bibfield  {author} {\bibinfo {author} {\bibfnamefont {L.}~\bibnamefont
    {Xiao}}, \bibinfo {author} {\bibfnamefont {G.~L.}\ \bibnamefont {Long}},
    \bibinfo {author} {\bibfnamefont {F.-G.}\ \bibnamefont {Deng}},\ and\
    \bibinfo {author} {\bibfnamefont {J.-W.}\ \bibnamefont {Pan}},\ }\bibfield
    {title} {\bibinfo {title} {Efficient multiparty quantum-secret-sharing
    schemes},\ }\href {https://doi.org/10.1103/PhysRevA.69.052307} {\bibfield
    {journal} {\bibinfo  {journal} {Physical Review A}\ }\textbf {\bibinfo
    {volume} {69}},\ \bibinfo {pages} {052307} (\bibinfo {year}
    {2004})}\BibitemShut {NoStop}%
  \bibitem [{\citenamefont {Long}\ and\ \citenamefont
    {Liu}(2002)}]{long2002theoretically}%
    \BibitemOpen
    \bibfield  {author} {\bibinfo {author} {\bibfnamefont {G.-L.}\ \bibnamefont
    {Long}}\ and\ \bibinfo {author} {\bibfnamefont {X.-S.}\ \bibnamefont {Liu}},\
    }\bibfield  {title} {\bibinfo {title} {Theoretically efficient high-capacity
    quantum-key-distribution scheme},\ }\href
    {https://doi.org/10.1103/PhysRevA.65.032302} {\bibfield  {journal} {\bibinfo
    {journal} {Physical Review A}\ }\textbf {\bibinfo {volume} {65}},\ \bibinfo
    {pages} {032302} (\bibinfo {year} {2002})}\BibitemShut {NoStop}%
  \bibitem [{\citenamefont {Deng}\ \emph {et~al.}(2003)\citenamefont {Deng},
    \citenamefont {Long},\ and\ \citenamefont {Liu}}]{deng2003two}%
    \BibitemOpen
    \bibfield  {author} {\bibinfo {author} {\bibfnamefont {F.-G.}\ \bibnamefont
    {Deng}}, \bibinfo {author} {\bibfnamefont {G.~L.}\ \bibnamefont {Long}},\
    and\ \bibinfo {author} {\bibfnamefont {X.-S.}\ \bibnamefont {Liu}},\
    }\bibfield  {title} {\bibinfo {title} {Two-step quantum direct communication
    protocol using the {Einstein-Podolsky-Rosen} pair block},\ }\href
    {https://doi.org/10.1103/PhysRevA.68.042317} {\bibfield  {journal} {\bibinfo
    {journal} {Physical Review A}\ }\textbf {\bibinfo {volume} {68}},\ \bibinfo
    {pages} {042317} (\bibinfo {year} {2003})}\BibitemShut {NoStop}%
  \bibitem [{\citenamefont {Shapiro}\ \emph {et~al.}(2019)\citenamefont
    {Shapiro}, \citenamefont {Boroson}, \citenamefont {Dixon}, \citenamefont
    {Grein},\ and\ \citenamefont {Hamilton}}]{shapiro2019quantum}%
    \BibitemOpen
    \bibfield  {author} {\bibinfo {author} {\bibfnamefont {J.~H.}\ \bibnamefont
    {Shapiro}}, \bibinfo {author} {\bibfnamefont {D.~M.}\ \bibnamefont
    {Boroson}}, \bibinfo {author} {\bibfnamefont {P.~B.}\ \bibnamefont {Dixon}},
    \bibinfo {author} {\bibfnamefont {M.~E.}\ \bibnamefont {Grein}},\ and\
    \bibinfo {author} {\bibfnamefont {S.~A.}\ \bibnamefont {Hamilton}},\
    }\bibfield  {title} {\bibinfo {title} {Quantum low probability of
    intercept},\ }\href {https://doi.org/10.1364/JOSAB.36.000B41} {\bibfield
    {journal} {\bibinfo  {journal} {JOSA B}\ }\textbf {\bibinfo {volume} {36}},\
    \bibinfo {pages} {B41} (\bibinfo {year} {2019})}\BibitemShut {NoStop}%
  \bibitem [{\citenamefont {Chandra}\ \emph {et~al.}(2021)\citenamefont
    {Chandra}, \citenamefont {Cacciapuoti}, \citenamefont {Caleffi},\ and\
    \citenamefont {Hanzo}}]{chandra2021direct}%
    \BibitemOpen
    \bibfield  {author} {\bibinfo {author} {\bibfnamefont {D.}~\bibnamefont
    {Chandra}}, \bibinfo {author} {\bibfnamefont {A.~S.}\ \bibnamefont
    {Cacciapuoti}}, \bibinfo {author} {\bibfnamefont {M.}~\bibnamefont
    {Caleffi}},\ and\ \bibinfo {author} {\bibfnamefont {L.}~\bibnamefont
    {Hanzo}},\ }\bibfield  {title} {\bibinfo {title} {Direct quantum
    communications in the presence of realistic noisy entanglement},\ }\href
    {https://doi.org/10.1109/TCOMM.2021.3122786} {\bibfield  {journal} {\bibinfo
    {journal} {IEEE Transactions on Communications}\ }\textbf {\bibinfo {volume}
    {70}},\ \bibinfo {pages} {469} (\bibinfo {year} {2021})}\BibitemShut
    {NoStop}%
  \bibitem [{\citenamefont {Wu}\ \emph {et~al.}(2022)\citenamefont {Wu},
    \citenamefont {Long},\ and\ \citenamefont {Hayashi}}]{wu2022quantum}%
    \BibitemOpen
    \bibfield  {author} {\bibinfo {author} {\bibfnamefont {J.}~\bibnamefont
    {Wu}}, \bibinfo {author} {\bibfnamefont {G.-L.}\ \bibnamefont {Long}},\ and\
    \bibinfo {author} {\bibfnamefont {M.}~\bibnamefont {Hayashi}},\ }\bibfield
    {title} {\bibinfo {title} {Quantum secure direct communication with private
    dense coding using a general preshared quantum state},\ }\href
    {https://doi.org/10.1103/PhysRevApplied.17.064011} {\bibfield  {journal}
    {\bibinfo  {journal} {Physical Review Applied}\ }\textbf {\bibinfo {volume}
    {17}},\ \bibinfo {pages} {064011} (\bibinfo {year} {2022})}\BibitemShut
    {NoStop}%
  \bibitem [{\citenamefont {Pan}\ \emph {et~al.}(2024)\citenamefont {Pan},
    \citenamefont {Long}, \citenamefont {Yin}, \citenamefont {Sheng},
    \citenamefont {Ruan}, \citenamefont {Ng}, \citenamefont {Lu},\ and\
    \citenamefont {Hanzo}}]{pan2023evolution}%
    \BibitemOpen
    \bibfield  {author} {\bibinfo {author} {\bibfnamefont {D.}~\bibnamefont
    {Pan}}, \bibinfo {author} {\bibfnamefont {G.-L.}\ \bibnamefont {Long}},
    \bibinfo {author} {\bibfnamefont {L.}~\bibnamefont {Yin}}, \bibinfo {author}
    {\bibfnamefont {Y.-B.}\ \bibnamefont {Sheng}}, \bibinfo {author}
    {\bibfnamefont {D.}~\bibnamefont {Ruan}}, \bibinfo {author} {\bibfnamefont
    {S.~X.}\ \bibnamefont {Ng}}, \bibinfo {author} {\bibfnamefont
    {J.}~\bibnamefont {Lu}},\ and\ \bibinfo {author} {\bibfnamefont
    {L.}~\bibnamefont {Hanzo}},\ }\bibfield  {title} {\bibinfo {title} {The
    evolution of quantum secure direct communication: On the road to the
    qinternet},\ }\href {https://doi.org/10.1109/COMST.2024.3367535} {\bibfield
    {journal} {\bibinfo  {journal} {IEEE Communications Surveys \& Tutorials}\ ,\
    \bibinfo {pages} {1}} (\bibinfo {year} {2024})}\BibitemShut {NoStop}%
  \bibitem [{\citenamefont {Sternberg}\ \emph {et~al.}(2024)\citenamefont
    {Sternberg}, \citenamefont {Voisin}, \citenamefont {Roux}, \citenamefont
    {Chassagneux},\ and\ \citenamefont {Amanti}}]{sternberg2024secure}%
    \BibitemOpen
    \bibfield  {author} {\bibinfo {author} {\bibfnamefont {J.}~\bibnamefont
    {Sternberg}}, \bibinfo {author} {\bibfnamefont {J.}~\bibnamefont {Voisin}},
    \bibinfo {author} {\bibfnamefont {C.}~\bibnamefont {Roux}}, \bibinfo {author}
    {\bibfnamefont {Y.}~\bibnamefont {Chassagneux}},\ and\ \bibinfo {author}
    {\bibfnamefont {M.~I.}\ \bibnamefont {Amanti}},\ }\bibfield  {title}
    {\bibinfo {title} {Secure communication based on sensing of undetected
    photons},\ }\href {https://arxiv.org/abs/2403.15557} {\bibfield  {journal}
    {\bibinfo  {journal} {arXiv preprint arXiv:2403.15557}\ } (\bibinfo {year}
    {2024})}\BibitemShut {NoStop}%
  \bibitem [{\citenamefont {Pan}\ \emph {et~al.}(2023)\citenamefont {Pan},
    \citenamefont {Song},\ and\ \citenamefont {Long}}]{pan2023free}%
    \BibitemOpen
    \bibfield  {author} {\bibinfo {author} {\bibfnamefont {D.}~\bibnamefont
    {Pan}}, \bibinfo {author} {\bibfnamefont {X.-T.}\ \bibnamefont {Song}},\ and\
    \bibinfo {author} {\bibfnamefont {G.-L.}\ \bibnamefont {Long}},\ }\bibfield
    {title} {\bibinfo {title} {Free-space quantum secure direct communication:
    Basics, progress, and outlook},\ }\href {https://doi.org/10.34133/adi.0004}
    {\bibfield  {journal} {\bibinfo  {journal} {Advanced Devices \&
    Instrumentation}\ }\textbf {\bibinfo {volume} {4}},\ \bibinfo {pages} {0004}
    (\bibinfo {year} {2023})}\BibitemShut {NoStop}%
  \bibitem [{\citenamefont {Cao}\ \emph {et~al.}(2023)\citenamefont {Cao},
    \citenamefont {Lu}, \citenamefont {Chai}, \citenamefont {Yu}, \citenamefont
    {Liang},\ and\ \citenamefont {Wang}}]{cao2023realization}%
    \BibitemOpen
    \bibfield  {author} {\bibinfo {author} {\bibfnamefont {Z.}~\bibnamefont
    {Cao}}, \bibinfo {author} {\bibfnamefont {Y.}~\bibnamefont {Lu}}, \bibinfo
    {author} {\bibfnamefont {G.}~\bibnamefont {Chai}}, \bibinfo {author}
    {\bibfnamefont {H.}~\bibnamefont {Yu}}, \bibinfo {author} {\bibfnamefont
    {K.}~\bibnamefont {Liang}},\ and\ \bibinfo {author} {\bibfnamefont
    {L.}~\bibnamefont {Wang}},\ }\bibfield  {title} {\bibinfo {title}
    {Realization of quantum secure direct communication with continuous
    variable},\ }\href {https://doi.org/10.34133/research.0193} {\bibfield
    {journal} {\bibinfo  {journal} {Research}\ }\textbf {\bibinfo {volume} {6}},\
    \bibinfo {pages} {0193} (\bibinfo {year} {2023})}\BibitemShut {NoStop}%
  \bibitem [{\citenamefont {Patra}\ \emph {et~al.}(2023)\citenamefont {Patra},
    \citenamefont {Gupta}, \citenamefont {Das},\ and\ \citenamefont
    {De}}]{patra2023dimensional}%
    \BibitemOpen
    \bibfield  {author} {\bibinfo {author} {\bibfnamefont {A.}~\bibnamefont
    {Patra}}, \bibinfo {author} {\bibfnamefont {R.}~\bibnamefont {Gupta}},
    \bibinfo {author} {\bibfnamefont {T.}~\bibnamefont {Das}},\ and\ \bibinfo
    {author} {\bibfnamefont {A.~S.}\ \bibnamefont {De}},\ }\bibfield  {title}
    {\bibinfo {title} {Dimensional advantage in secure information trading via
    the noisy dense coding protocol},\ }\href {https://arxiv.org/abs/2310.20688}
    {\bibfield  {journal} {\bibinfo  {journal} {arXiv preprint arXiv:2310.20688}\
    } (\bibinfo {year} {2023})}\BibitemShut {NoStop}%
  \bibitem [{\citenamefont {Sephton}\ \emph {et~al.}(2023)\citenamefont
    {Sephton}, \citenamefont {Vall{\'e}s}, \citenamefont {Nape}, \citenamefont
    {Cox}, \citenamefont {Steinlechner}, \citenamefont {Konrad}, \citenamefont
    {Torres}, \citenamefont {Roux},\ and\ \citenamefont
    {Forbes}}]{sephton2023quantum}%
    \BibitemOpen
    \bibfield  {author} {\bibinfo {author} {\bibfnamefont {B.}~\bibnamefont
    {Sephton}}, \bibinfo {author} {\bibfnamefont {A.}~\bibnamefont {Vall{\'e}s}},
    \bibinfo {author} {\bibfnamefont {I.}~\bibnamefont {Nape}}, \bibinfo {author}
    {\bibfnamefont {M.~A.}\ \bibnamefont {Cox}}, \bibinfo {author} {\bibfnamefont
    {F.}~\bibnamefont {Steinlechner}}, \bibinfo {author} {\bibfnamefont
    {T.}~\bibnamefont {Konrad}}, \bibinfo {author} {\bibfnamefont {J.~P.}\
    \bibnamefont {Torres}}, \bibinfo {author} {\bibfnamefont {F.~S.}\
    \bibnamefont {Roux}},\ and\ \bibinfo {author} {\bibfnamefont
    {A.}~\bibnamefont {Forbes}},\ }\bibfield  {title} {\bibinfo {title} {Quantum
    transport of high-dimensional spatial information with a nonlinear
    detector},\ }\href {https://doi.org/10.1038/s41467-023-43949-x} {\bibfield
    {journal} {\bibinfo  {journal} {Nature Communications}\ }\textbf {\bibinfo
    {volume} {14}},\ \bibinfo {pages} {8243} (\bibinfo {year}
    {2023})}\BibitemShut {NoStop}%
  \bibitem [{\citenamefont {Zhou}\ \emph {et~al.}(2020)\citenamefont {Zhou},
    \citenamefont {Sheng},\ and\ \citenamefont {Long}}]{zhou2020device}%
    \BibitemOpen
    \bibfield  {author} {\bibinfo {author} {\bibfnamefont {L.}~\bibnamefont
    {Zhou}}, \bibinfo {author} {\bibfnamefont {Y.-B.}\ \bibnamefont {Sheng}},\
    and\ \bibinfo {author} {\bibfnamefont {G.-L.}\ \bibnamefont {Long}},\
    }\bibfield  {title} {\bibinfo {title} {Device-independent quantum secure
    direct communication against collective attacks},\ }\href
    {https://doi.org/https://doi.org/10.1016/j.scib.2019.10.025} {\bibfield
    {journal} {\bibinfo  {journal} {Science Bulletin}\ }\textbf {\bibinfo
    {volume} {65}},\ \bibinfo {pages} {12} (\bibinfo {year} {2020})}\BibitemShut
    {NoStop}%
  \bibitem [{\citenamefont {Zhang}\ \emph {et~al.}(2017)\citenamefont {Zhang},
    \citenamefont {Ding}, \citenamefont {Sheng}, \citenamefont {Zhou},
    \citenamefont {Shi},\ and\ \citenamefont {Guo}}]{zhang2017quantum}%
    \BibitemOpen
    \bibfield  {author} {\bibinfo {author} {\bibfnamefont {W.}~\bibnamefont
    {Zhang}}, \bibinfo {author} {\bibfnamefont {D.-S.}\ \bibnamefont {Ding}},
    \bibinfo {author} {\bibfnamefont {Y.-B.}\ \bibnamefont {Sheng}}, \bibinfo
    {author} {\bibfnamefont {L.}~\bibnamefont {Zhou}}, \bibinfo {author}
    {\bibfnamefont {B.-S.}\ \bibnamefont {Shi}},\ and\ \bibinfo {author}
    {\bibfnamefont {G.-C.}\ \bibnamefont {Guo}},\ }\bibfield  {title} {\bibinfo
    {title} {Quantum secure direct communication with quantum memory},\ }\href
    {https://doi.org/10.1103/PhysRevLett.118.220501} {\bibfield  {journal}
    {\bibinfo  {journal} {Physical Review Letters}\ }\textbf {\bibinfo {volume}
    {118}},\ \bibinfo {pages} {220501} (\bibinfo {year} {2017})}\BibitemShut
    {NoStop}%
  \bibitem [{\citenamefont {Zhu}\ \emph {et~al.}(2017)\citenamefont {Zhu},
    \citenamefont {Zhang}, \citenamefont {Sheng},\ and\ \citenamefont
    {Huang}}]{zhu2017experimental}%
    \BibitemOpen
    \bibfield  {author} {\bibinfo {author} {\bibfnamefont {F.}~\bibnamefont
    {Zhu}}, \bibinfo {author} {\bibfnamefont {W.}~\bibnamefont {Zhang}}, \bibinfo
    {author} {\bibfnamefont {Y.}~\bibnamefont {Sheng}},\ and\ \bibinfo {author}
    {\bibfnamefont {Y.}~\bibnamefont {Huang}},\ }\bibfield  {title} {\bibinfo
    {title} {Experimental long-distance quantum secure direct communication},\
    }\href {https://doi.org/https://doi.org/10.1016/j.scib.2017.10.023}
    {\bibfield  {journal} {\bibinfo  {journal} {Science Bulletin}\ }\textbf
    {\bibinfo {volume} {62}},\ \bibinfo {pages} {1519} (\bibinfo {year}
    {2017})}\BibitemShut {NoStop}%
    \bibitem [{\citenamefont {Nirala}\ \emph {et~al.}(2023)\citenamefont {Nirala},
    \citenamefont {Pradyumna}, \citenamefont {Kumar},\ and\ \citenamefont
    {Marino}}]{nirala2023information}%
    \BibitemOpen
    \bibfield  {author} {\bibinfo {author} {\bibfnamefont {G.}~\bibnamefont
    {Nirala}}, \bibinfo {author} {\bibfnamefont {S.~T.}\ \bibnamefont
    {Pradyumna}}, \bibinfo {author} {\bibfnamefont {A.}~\bibnamefont {Kumar}},\
    and\ \bibinfo {author} {\bibfnamefont {A.~M.}\ \bibnamefont {Marino}},\
    }\bibfield  {title} {\bibinfo {title} {Information encoding in the spatial
    correlations of entangled twin beams},\ }\href
    {https://doi.org/10.1126/sciadv.adf9161} {\bibfield  {journal} {\bibinfo
    {journal} {Science Advances}\ }\textbf {\bibinfo {volume} {9}},\ \bibinfo
    {pages} {eadf9161} (\bibinfo {year} {2023})}\BibitemShut {NoStop}%
  \bibitem [{\citenamefont {Qi}\ \emph {et~al.}(2021)\citenamefont {Qi},
    \citenamefont {Li}, \citenamefont {Huang}, \citenamefont {Feng},
    \citenamefont {Zheng},\ and\ \citenamefont {Chen}}]{qi202115}%
    \BibitemOpen
    \bibfield  {author} {\bibinfo {author} {\bibfnamefont {Z.}~\bibnamefont
    {Qi}}, \bibinfo {author} {\bibfnamefont {Y.}~\bibnamefont {Li}}, \bibinfo
    {author} {\bibfnamefont {Y.}~\bibnamefont {Huang}}, \bibinfo {author}
    {\bibfnamefont {J.}~\bibnamefont {Feng}}, \bibinfo {author} {\bibfnamefont
    {Y.}~\bibnamefont {Zheng}},\ and\ \bibinfo {author} {\bibfnamefont
    {X.}~\bibnamefont {Chen}},\ }\bibfield  {title} {\bibinfo {title} {A 15-user
    quantum secure direct communication network},\ }\href
    {https://doi.org/10.1038/s41377-021-00634-2} {\bibfield  {journal} {\bibinfo
    {journal} {Light: Science \& Applications}\ }\textbf {\bibinfo {volume}
    {10}},\ \bibinfo {pages} {183} (\bibinfo {year} {2021})}\BibitemShut
    {NoStop}%
    \bibitem [{\citenamefont {Degen}\ \emph {et~al.}(2017)\citenamefont {Degen},
    \citenamefont {Reinhard},\ and\ \citenamefont
    {Cappellaro}}]{degen2017quantum}%
    \BibitemOpen
    \bibfield  {author} {\bibinfo {author} {\bibfnamefont {C.~L.}\ \bibnamefont
    {Degen}}, \bibinfo {author} {\bibfnamefont {F.}~\bibnamefont {Reinhard}},\
    and\ \bibinfo {author} {\bibfnamefont {P.}~\bibnamefont {Cappellaro}},\
    }\bibfield  {title} {\bibinfo {title} {Quantum sensing},\ }\href
    {https://doi.org/10.1103/RevModPhys.89.035002} {\bibfield  {journal}
    {\bibinfo  {journal} {Reviews of Modern Physics}\ }\textbf {\bibinfo {volume}
    {89}},\ \bibinfo {pages} {035002} (\bibinfo {year} {2017})}\BibitemShut
    {NoStop}%
  \bibitem [{\citenamefont {Giovannetti}\ \emph {et~al.}(2006)\citenamefont
    {Giovannetti}, \citenamefont {Lloyd},\ and\ \citenamefont
    {Maccone}}]{giovannetti2006quantum}%
    \BibitemOpen
    \bibfield  {author} {\bibinfo {author} {\bibfnamefont {V.}~\bibnamefont
    {Giovannetti}}, \bibinfo {author} {\bibfnamefont {S.}~\bibnamefont {Lloyd}},\
    and\ \bibinfo {author} {\bibfnamefont {L.}~\bibnamefont {Maccone}},\
    }\bibfield  {title} {\bibinfo {title} {Quantum metrology},\ }\href
    {https://doi.org/10.1103/PhysRevLett.96.010401} {\bibfield  {journal}
    {\bibinfo  {journal} {Physical Review letters}\ }\textbf {\bibinfo {volume}
    {96}},\ \bibinfo {pages} {010401} (\bibinfo {year} {2006})}\BibitemShut
    {NoStop}%
  \bibitem [{\citenamefont {Giovannetti}\ \emph {et~al.}(2011)\citenamefont
    {Giovannetti}, \citenamefont {Lloyd},\ and\ \citenamefont
    {Maccone}}]{giovannetti2011advances}%
    \BibitemOpen
    \bibfield  {author} {\bibinfo {author} {\bibfnamefont {V.}~\bibnamefont
    {Giovannetti}}, \bibinfo {author} {\bibfnamefont {S.}~\bibnamefont {Lloyd}},\
    and\ \bibinfo {author} {\bibfnamefont {L.}~\bibnamefont {Maccone}},\
    }\bibfield  {title} {\bibinfo {title} {Advances in quantum metrology},\
    }\href {https://doi.org/10.1038/nphoton.2011.35} {\bibfield  {journal}
    {\bibinfo  {journal} {Nature Photonics}\ }\textbf {\bibinfo {volume} {5}},\
    \bibinfo {pages} {222} (\bibinfo {year} {2011})}\BibitemShut {NoStop}%
  \bibitem [{\citenamefont {T{\'o}th}\ and\ \citenamefont
    {Apellaniz}(2014)}]{toth2014quantum}%
    \BibitemOpen
    \bibfield  {author} {\bibinfo {author} {\bibfnamefont {G.}~\bibnamefont
    {T{\'o}th}}\ and\ \bibinfo {author} {\bibfnamefont {I.}~\bibnamefont
    {Apellaniz}},\ }\bibfield  {title} {\bibinfo {title} {Quantum metrology from
    a quantum information science perspective},\ }\href
    {https://doi.org/10.1088/1751-8113/47/42/424006} {\bibfield  {journal}
    {\bibinfo  {journal} {Journal of Physics A: Mathematical and Theoretical}\
    }\textbf {\bibinfo {volume} {47}},\ \bibinfo {pages} {424006} (\bibinfo
    {year} {2014})}\BibitemShut {NoStop}%
  \bibitem [{\citenamefont {Pezze}\ \emph {et~al.}(2018)\citenamefont {Pezze},
    \citenamefont {Smerzi}, \citenamefont {Oberthaler}, \citenamefont {Schmied},\
    and\ \citenamefont {Treutlein}}]{pezze2018quantum}%
    \BibitemOpen
    \bibfield  {author} {\bibinfo {author} {\bibfnamefont {L.}~\bibnamefont
    {Pezze}}, \bibinfo {author} {\bibfnamefont {A.}~\bibnamefont {Smerzi}},
    \bibinfo {author} {\bibfnamefont {M.~K.}\ \bibnamefont {Oberthaler}},
    \bibinfo {author} {\bibfnamefont {R.}~\bibnamefont {Schmied}},\ and\ \bibinfo
    {author} {\bibfnamefont {P.}~\bibnamefont {Treutlein}},\ }\bibfield  {title}
    {\bibinfo {title} {Quantum metrology with nonclassical states of atomic
    ensembles},\ }\href {https://doi.org/10.1103/RevModPhys.90.035005} {\bibfield
     {journal} {\bibinfo  {journal} {Reviews of Modern Physics}\ }\textbf
    {\bibinfo {volume} {90}},\ \bibinfo {pages} {035005} (\bibinfo {year}
    {2018})}\BibitemShut {NoStop}%
  \bibitem [{\citenamefont {Moore}\ and\ \citenamefont
    {Dunningham}(2023)}]{moore2023secure}%
    \BibitemOpen
    \bibfield  {author} {\bibinfo {author} {\bibfnamefont {S.~W.}\ \bibnamefont
    {Moore}}\ and\ \bibinfo {author} {\bibfnamefont {J.~A.}\ \bibnamefont
    {Dunningham}},\ }\bibfield  {title} {\bibinfo {title} {Secure quantum remote
    sensing without entanglement},\ }\bibfield  {journal} {\bibinfo  {journal}
    {AVS Quantum Science}\ }\textbf {\bibinfo {volume} {5}},\ \href
    {https://doi.org/10.1116/5.0137260} {10.1116/5.0137260} (\bibinfo {year}
    {2023})\BibitemShut {NoStop}%
  \bibitem [{\citenamefont {Zhang}\ \emph {et~al.}(2021)\citenamefont {Zhang},
    \citenamefont {Qin}, \citenamefont {Song},\ and\ \citenamefont
    {Long}}]{zhang2021color}%
    \BibitemOpen
    \bibfield  {author} {\bibinfo {author} {\bibfnamefont {H.}~\bibnamefont
    {Zhang}}, \bibinfo {author} {\bibfnamefont {G.-Q.}\ \bibnamefont {Qin}},
    \bibinfo {author} {\bibfnamefont {X.-K.}\ \bibnamefont {Song}},\ and\
    \bibinfo {author} {\bibfnamefont {G.-L.}\ \bibnamefont {Long}},\ }\bibfield
    {title} {\bibinfo {title} {Color-detuning-dynamics-based quantum sensing with
    dressed states driving},\ }\href {https://doi.org/10.1364/OE.413637}
    {\bibfield  {journal} {\bibinfo  {journal} {Optics Express}\ }\textbf
    {\bibinfo {volume} {29}},\ \bibinfo {pages} {5358} (\bibinfo {year}
    {2021})}\BibitemShut {NoStop}%
    \bibitem [{\citenamefont {Ramsey}(1950)}]{Ramsey1950A}%
    \BibitemOpen
    \bibfield  {author} {\bibinfo {author} {\bibfnamefont {N.~F.}\ \bibnamefont
    {Ramsey}},\ }\bibfield  {title} {\bibinfo {title} {A molecular beam resonance
    method with separated oscillating fields},\ }\href
    {https://doi.org/10.1103/PhysRev.78.695} {\bibfield  {journal} {\bibinfo
    {journal} {Phys. Rev.}\ }\textbf {\bibinfo {volume} {78}},\ \bibinfo {pages}
    {695} (\bibinfo {year} {1950})}\BibitemShut {NoStop}%
  \bibitem [{\citenamefont {Lau}\ \emph {et~al.}(2022)\citenamefont {Lau},
    \citenamefont {Lim}, \citenamefont {Shrotriya},\ and\ \citenamefont
    {Kwek}}]{lau2022nisq}%
    \BibitemOpen
    \bibfield  {author} {\bibinfo {author} {\bibfnamefont {J.~W.~Z.}\
    \bibnamefont {Lau}}, \bibinfo {author} {\bibfnamefont {K.~H.}\ \bibnamefont
    {Lim}}, \bibinfo {author} {\bibfnamefont {H.}~\bibnamefont {Shrotriya}},\
    and\ \bibinfo {author} {\bibfnamefont {L.~C.}\ \bibnamefont {Kwek}},\
    }\bibfield  {title} {\bibinfo {title} {{NISQ} computing: where are we and
    where do we go?},\ }\href {https://doi.org/10.1007/s43673-022-00058-z}
    {\bibfield  {journal} {\bibinfo  {journal} {AAPPS bulletin}\ }\textbf
    {\bibinfo {volume} {32}},\ \bibinfo {pages} {27} (\bibinfo {year}
    {2022})}\BibitemShut {NoStop}%
  \bibitem [{\citenamefont {Yin}\ \emph {et~al.}(2020)\citenamefont {Yin},
    \citenamefont {Takeuchi}, \citenamefont {Zhang}, \citenamefont {Yin},
    \citenamefont {Matsuzaki}, \citenamefont {Peng}, \citenamefont {Xu},
    \citenamefont {Xu}, \citenamefont {Tang}, \citenamefont {Zhou} \emph
    {et~al.}}]{yin2020experimental}%
    \BibitemOpen
    \bibfield  {author} {\bibinfo {author} {\bibfnamefont {P.}~\bibnamefont
    {Yin}}, \bibinfo {author} {\bibfnamefont {Y.}~\bibnamefont {Takeuchi}},
    \bibinfo {author} {\bibfnamefont {W.-H.}\ \bibnamefont {Zhang}}, \bibinfo
    {author} {\bibfnamefont {Z.-Q.}\ \bibnamefont {Yin}}, \bibinfo {author}
    {\bibfnamefont {Y.}~\bibnamefont {Matsuzaki}}, \bibinfo {author}
    {\bibfnamefont {X.-X.}\ \bibnamefont {Peng}}, \bibinfo {author}
    {\bibfnamefont {X.-Y.}\ \bibnamefont {Xu}}, \bibinfo {author} {\bibfnamefont
    {J.-S.}\ \bibnamefont {Xu}}, \bibinfo {author} {\bibfnamefont {J.-S.}\
    \bibnamefont {Tang}}, \bibinfo {author} {\bibfnamefont {Z.-Q.}\ \bibnamefont
    {Zhou}}, \emph {et~al.},\ }\bibfield  {title} {\bibinfo {title} {Experimental
    demonstration of secure quantum remote sensing},\ }\href
    {https://doi.org/10.1103/PhysRevApplied.14.014065} {\bibfield  {journal}
    {\bibinfo  {journal} {Physical Review Applied}\ }\textbf {\bibinfo {volume}
    {14}},\ \bibinfo {pages} {014065} (\bibinfo {year} {2020})}\BibitemShut
    {NoStop}%
  \bibitem [{\citenamefont {Rahim}\ \emph {et~al.}(2023)\citenamefont {Rahim},
    \citenamefont {Khan}, \citenamefont {Khalid}, \citenamefont {Rehman},
    \citenamefont {Jung},\ and\ \citenamefont {Shin}}]{rahim2023quantum}%
    \BibitemOpen
    \bibfield  {author} {\bibinfo {author} {\bibfnamefont {M.~T.}\ \bibnamefont
    {Rahim}}, \bibinfo {author} {\bibfnamefont {A.}~\bibnamefont {Khan}},
    \bibinfo {author} {\bibfnamefont {U.}~\bibnamefont {Khalid}}, \bibinfo
    {author} {\bibfnamefont {J.~u.}\ \bibnamefont {Rehman}}, \bibinfo {author}
    {\bibfnamefont {H.}~\bibnamefont {Jung}},\ and\ \bibinfo {author}
    {\bibfnamefont {H.}~\bibnamefont {Shin}},\ }\bibfield  {title} {\bibinfo
    {title} {Quantum secure metrology for network sensing-based applications},\
    }\href {https://doi.org/10.1038/s41598-023-38802-6} {\bibfield  {journal}
    {\bibinfo  {journal} {Scientific Reports}\ }\textbf {\bibinfo {volume}
    {13}},\ \bibinfo {pages} {11630} (\bibinfo {year} {2023})}\BibitemShut
    {NoStop}%
  \bibitem [{\citenamefont {Fahmi}\ and\ \citenamefont
    {Golshani}(2008)}]{fahmi2008comment}%
    \BibitemOpen
    \bibfield  {author} {\bibinfo {author} {\bibfnamefont {A.}~\bibnamefont
    {Fahmi}}\ and\ \bibinfo {author} {\bibfnamefont {M.}~\bibnamefont
    {Golshani}},\ }\bibfield  {title} {\bibinfo {title} {Comment on “quantum
    key distribution in the holevo limit”},\ }\href
    {https://doi.org/10.1103/PhysRevLett.100.018901} {\bibfield  {journal}
    {\bibinfo  {journal} {Physical Review Letters}\ }\textbf {\bibinfo {volume}
    {100}},\ \bibinfo {pages} {018901} (\bibinfo {year} {2008})}\BibitemShut
    {NoStop}%
    \bibitem [{\citenamefont {Beaudry}\ \emph {et~al.}(2013)\citenamefont
    {Beaudry}, \citenamefont {Lucamarini}, \citenamefont {Mancini},\ and\
    \citenamefont {Renner}}]{beaudry2013security}%
    \BibitemOpen
    \bibfield  {author} {\bibinfo {author} {\bibfnamefont {N.~J.}\ \bibnamefont
    {Beaudry}}, \bibinfo {author} {\bibfnamefont {M.}~\bibnamefont {Lucamarini}},
    \bibinfo {author} {\bibfnamefont {S.}~\bibnamefont {Mancini}},\ and\ \bibinfo
    {author} {\bibfnamefont {R.}~\bibnamefont {Renner}},\ }\bibfield  {title}
    {\bibinfo {title} {Security of two-way quantum key distribution},\ }\href
    {https://doi.org/https://doi.org/10.1103/PhysRevA.88.062302} {\bibfield
    {journal} {\bibinfo  {journal} {Physical Review A}\ }\textbf {\bibinfo
    {volume} {88}},\ \bibinfo {pages} {062302} (\bibinfo {year}
    {2013})}\BibitemShut {NoStop}%
  \bibitem [{\citenamefont {Wu}\ \emph {et~al.}(2019)\citenamefont {Wu},
    \citenamefont {Lin}, \citenamefont {Yin},\ and\ \citenamefont
    {Long}}]{wu2019security}%
    \BibitemOpen
    \bibfield  {author} {\bibinfo {author} {\bibfnamefont {J.}~\bibnamefont
    {Wu}}, \bibinfo {author} {\bibfnamefont {Z.}~\bibnamefont {Lin}}, \bibinfo
    {author} {\bibfnamefont {L.}~\bibnamefont {Yin}},\ and\ \bibinfo {author}
    {\bibfnamefont {G.-L.}\ \bibnamefont {Long}},\ }\bibfield  {title} {\bibinfo
    {title} {Security of quantum secure direct communication based on {Wyner's}
    wiretap channel theory},\ }\href
    {https://doi.org/https://doi.org/10.1002/que2.26} {\bibfield  {journal}
    {\bibinfo  {journal} {Quantum Engineering}\ }\textbf {\bibinfo {volume}
    {1}},\ \bibinfo {pages} {e26} (\bibinfo {year} {2019})}\BibitemShut {NoStop}%
  \bibitem [{\citenamefont {Zhong}\ \emph {et~al.}(2014)\citenamefont {Zhong},
    \citenamefont {Lu}, \citenamefont {Jing},\ and\ \citenamefont
    {Wang}}]{zhong2014optimal}%
    \BibitemOpen
    \bibfield  {author} {\bibinfo {author} {\bibfnamefont {W.}~\bibnamefont
    {Zhong}}, \bibinfo {author} {\bibfnamefont {X.~M.}\ \bibnamefont {Lu}},
    \bibinfo {author} {\bibfnamefont {X.~X.}\ \bibnamefont {Jing}},\ and\
    \bibinfo {author} {\bibfnamefont {X.}~\bibnamefont {Wang}},\ }\bibfield
    {title} {\bibinfo {title} {Optimal condition for measurement observable via
    error-propagation},\ }\href {https://doi.org/10.1088/1751-8113/47/38/385304}
    {\bibfield  {journal} {\bibinfo  {journal} {Journal of Physics A:
    Mathematical and Theoretical}\ }\textbf {\bibinfo {volume} {47}},\ \bibinfo
    {pages} {385304} (\bibinfo {year} {2014})}\BibitemShut {NoStop}%
  \bibitem [{\citenamefont {Braunstein}\ \emph {et~al.}(1996)\citenamefont
    {Braunstein}, \citenamefont {Caves},\ and\ \citenamefont
    {Milburn}}]{braunstein1996generalized}%
    \BibitemOpen
    \bibfield  {author} {\bibinfo {author} {\bibfnamefont {S.~L.}\ \bibnamefont
    {Braunstein}}, \bibinfo {author} {\bibfnamefont {C.~M.}\ \bibnamefont
    {Caves}},\ and\ \bibinfo {author} {\bibfnamefont {G.~J.}\ \bibnamefont
    {Milburn}},\ }\bibfield  {title} {\bibinfo {title} {Generalized uncertainty
    relations: theory, examples, and lorentz invariance},\ }\href
    {https://doi.org/https://doi.org/10.1006/aphy.1996.0040} {\bibfield
    {journal} {\bibinfo  {journal} {Annals of Physics}\ }\textbf {\bibinfo
    {volume} {247}},\ \bibinfo {pages} {135} (\bibinfo {year}
    {1996})}\BibitemShut {NoStop}%
  \bibitem [{\citenamefont {Helstrom}(1969)}]{helstrom1969quantum}%
    \BibitemOpen
    \bibfield  {author} {\bibinfo {author} {\bibfnamefont {C.~W.}\ \bibnamefont
    {Helstrom}},\ }\bibfield  {title} {\bibinfo {title} {Quantum detection and
    estimation theory},\ }\href
    {https://link.springer.com/article/10.1007/bf01007479} {\bibfield  {journal}
    {\bibinfo  {journal} {Journal of Statistical Physics}\ }\textbf {\bibinfo
    {volume} {1}},\ \bibinfo {pages} {231} (\bibinfo {year} {1969})}\BibitemShut
    {NoStop}%
  \bibitem [{\citenamefont {Liu}\ \emph {et~al.}(2014)\citenamefont {Liu},
    \citenamefont {Jing}, \citenamefont {Zhong},\ and\ \citenamefont
    {Wang}}]{liu2014quantum}%
    \BibitemOpen
    \bibfield  {author} {\bibinfo {author} {\bibfnamefont {J.}~\bibnamefont
    {Liu}}, \bibinfo {author} {\bibfnamefont {X.-X.}\ \bibnamefont {Jing}},
    \bibinfo {author} {\bibfnamefont {W.}~\bibnamefont {Zhong}},\ and\ \bibinfo
    {author} {\bibfnamefont {X.-G.}\ \bibnamefont {Wang}},\ }\bibfield  {title}
    {\bibinfo {title} {Quantum fisher information for density matrices with
    arbitrary ranks},\ }\href {https://doi.org/10.1088/0253-6102/61/1/08}
    {\bibfield  {journal} {\bibinfo  {journal} {Communications in Theoretical
    Physics}\ }\textbf {\bibinfo {volume} {61}},\ \bibinfo {pages} {45} (\bibinfo
    {year} {2014})}\BibitemShut {NoStop}%
    \bibitem [{\citenamefont {Braun}\ \emph {et~al.}(2018)\citenamefont {Braun},
    \citenamefont {Adesso}, \citenamefont {Benatti}, \citenamefont {Floreanini},
    \citenamefont {Marzolino}, \citenamefont {Mitchell},\ and\ \citenamefont
    {Pirandola}}]{braun2018quantum}%
    \BibitemOpen
    \bibfield  {author} {\bibinfo {author} {\bibfnamefont {D.}~\bibnamefont
    {Braun}}, \bibinfo {author} {\bibfnamefont {G.}~\bibnamefont {Adesso}},
    \bibinfo {author} {\bibfnamefont {F.}~\bibnamefont {Benatti}}, \bibinfo
    {author} {\bibfnamefont {R.}~\bibnamefont {Floreanini}}, \bibinfo {author}
    {\bibfnamefont {U.}~\bibnamefont {Marzolino}}, \bibinfo {author}
    {\bibfnamefont {M.~W.}\ \bibnamefont {Mitchell}},\ and\ \bibinfo {author}
    {\bibfnamefont {S.}~\bibnamefont {Pirandola}},\ }\bibfield  {title} {\bibinfo
    {title} {Quantum-enhanced measurements without entanglement},\ }\href
    {https://doi.org/10.1103/RevModPhys.90.035006} {\bibfield  {journal}
    {\bibinfo  {journal} {Reviews of Modern Physics}\ }\textbf {\bibinfo {volume}
    {90}},\ \bibinfo {pages} {035006} (\bibinfo {year} {2018})}\BibitemShut
    {NoStop}%
  \bibitem [{\citenamefont {Berry}\ \emph {et~al.}(2009)\citenamefont {Berry},
    \citenamefont {Higgins}, \citenamefont {Bartlett}, \citenamefont {Mitchell},
    \citenamefont {Pryde},\ and\ \citenamefont {Wiseman}}]{berry2009perform}%
    \BibitemOpen
    \bibfield  {author} {\bibinfo {author} {\bibfnamefont {D.~W.}\ \bibnamefont
    {Berry}}, \bibinfo {author} {\bibfnamefont {B.~L.}\ \bibnamefont {Higgins}},
    \bibinfo {author} {\bibfnamefont {S.~D.}\ \bibnamefont {Bartlett}}, \bibinfo
    {author} {\bibfnamefont {M.~W.}\ \bibnamefont {Mitchell}}, \bibinfo {author}
    {\bibfnamefont {G.~J.}\ \bibnamefont {Pryde}},\ and\ \bibinfo {author}
    {\bibfnamefont {H.~M.}\ \bibnamefont {Wiseman}},\ }\bibfield  {title}
    {\bibinfo {title} {How to perform the most accurate possible phase
    measurements},\ }\href {https://doi.org/10.1103/PhysRevA.80.052114}
    {\bibfield  {journal} {\bibinfo  {journal} {Physical Review A}\ }\textbf
    {\bibinfo {volume} {80}},\ \bibinfo {pages} {052114} (\bibinfo {year}
    {2009})}\BibitemShut {NoStop}%
  \bibitem [{\citenamefont {Renner}(2007)}]{renner2007symmetry}%
    \BibitemOpen
    \bibfield  {author} {\bibinfo {author} {\bibfnamefont {R.}~\bibnamefont
    {Renner}},\ }\bibfield  {title} {\bibinfo {title} {Symmetry of large physical
    systems implies independence of subsystems},\ }\href
    {https://doi.org/10.1038/nphys684} {\bibfield  {journal} {\bibinfo  {journal}
    {Nature Physics}\ }\textbf {\bibinfo {volume} {3}},\ \bibinfo {pages} {645}
    (\bibinfo {year} {2007})}\BibitemShut {NoStop}%
  \bibitem [{\citenamefont {Wyner}(1975)}]{wyner1975wire}%
    \BibitemOpen
    \bibfield  {author} {\bibinfo {author} {\bibfnamefont {A.~D.}\ \bibnamefont
    {Wyner}},\ }\bibfield  {title} {\bibinfo {title} {The wire-tap channel},\
    }\href {https://doi.org/https://doi.org/10.1002/j.1538-7305.1975.tb02040.x}
    {\bibfield  {journal} {\bibinfo  {journal} {Bell System Technical Journal}\
    }\textbf {\bibinfo {volume} {54}},\ \bibinfo {pages} {1355} (\bibinfo {year}
    {1975})}\BibitemShut {NoStop}%
  \bibitem [{\citenamefont {Kraus}\ \emph {et~al.}(2005)\citenamefont {Kraus},
    \citenamefont {Gisin},\ and\ \citenamefont {Renner}}]{kraus2005lower}%
    \BibitemOpen
    \bibfield  {author} {\bibinfo {author} {\bibfnamefont {B.}~\bibnamefont
    {Kraus}}, \bibinfo {author} {\bibfnamefont {N.}~\bibnamefont {Gisin}},\ and\
    \bibinfo {author} {\bibfnamefont {R.}~\bibnamefont {Renner}},\ }\bibfield
    {title} {\bibinfo {title} {Lower and upper bounds on the secret-key rate for
    quantum key distribution protocols using one-way classical communication},\
    }\href {https://doi.org/10.1103/PhysRevLett.95.080501} {\bibfield  {journal}
    {\bibinfo  {journal} {Physical Review Letters}\ }\textbf {\bibinfo {volume}
    {95}},\ \bibinfo {pages} {080501} (\bibinfo {year} {2005})}\BibitemShut
    {NoStop}%
    \bibitem [{\citenamefont {Zheng}\ \emph {et~al.}(2019)\citenamefont {Zheng},
    \citenamefont {Xu}, \citenamefont {Zhang}, \citenamefont {Ning},\ and\
    \citenamefont {Zhang}}]{zheng2019ab}%
    \BibitemOpen
    \bibfield  {author} {\bibinfo {author} {\bibfnamefont {K.}~\bibnamefont
    {Zheng}}, \bibinfo {author} {\bibfnamefont {H.}~\bibnamefont {Xu}}, \bibinfo
    {author} {\bibfnamefont {A.}~\bibnamefont {Zhang}}, \bibinfo {author}
    {\bibfnamefont {X.}~\bibnamefont {Ning}},\ and\ \bibinfo {author}
    {\bibfnamefont {L.}~\bibnamefont {Zhang}},\ }\bibfield  {title} {\bibinfo
    {title} {Ab initio phase estimation at the shot noise limit with on--off
    measurement},\ }\href {https://doi.org/10.1007/s11128-019-2450-z} {\bibfield
    {journal} {\bibinfo  {journal} {Quantum Information Processing}\ }\textbf
    {\bibinfo {volume} {18}},\ \bibinfo {pages} {1} (\bibinfo {year}
    {2019})}\BibitemShut {NoStop}%
    \bibitem [{\citenamefont {Wehner}\ \emph {et~al.}(2018)\citenamefont {Wehner},
    \citenamefont {Elkouss},\ and\ \citenamefont {Hanson}}]{wehner2018quantum}%
    \BibitemOpen
    \bibfield  {author} {\bibinfo {author} {\bibfnamefont {S.}~\bibnamefont
    {Wehner}}, \bibinfo {author} {\bibfnamefont {D.}~\bibnamefont {Elkouss}},\
    and\ \bibinfo {author} {\bibfnamefont {R.}~\bibnamefont {Hanson}},\
    }\bibfield  {title} {\bibinfo {title} {Quantum internet: A vision for the
    road ahead},\ }\href {https://doi.org/10.1126/science.aam9288} {\bibfield
    {journal} {\bibinfo  {journal} {Science}\ }\textbf {\bibinfo {volume}
    {362}},\ \bibinfo {pages} {eaam9288} (\bibinfo {year} {2018})}\BibitemShut
    {NoStop}
  \end{thebibliography}
%
  
\end{document}